\documentclass[lettersize,journal]{IEEEtran}
\IEEEoverridecommandlockouts
\usepackage{amsmath,amsfonts}
\usepackage{hyperref}
\usepackage{cite}
\hypersetup{colorlinks=true,citecolor=blue}
\usepackage{graphicx}
\usepackage{array}
\usepackage{booktabs} % For formal tables
\usepackage{multirow}
\usepackage{siunitx} % For table spacing
\usepackage{textcomp}
\usepackage{stfloats}
\usepackage{url}
\usepackage{booktabs} % For better looking tables
\usepackage{siunitx}  % Allows for alignment and formatting of numbers in tables
\usepackage{multirow} % Allows for multirow cells in tables
\usepackage{color}    % To use color
\usepackage{verbatim}
\usepackage{graphicx}
\usepackage{graphicx}
\usepackage{enumitem,kantlipsum}
\usepackage{subcaption}
\usepackage{tablefootnote}
\usepackage{cleveref}
\usepackage{svg}
\usepackage{textcomp}
\usepackage{breqn}
\usepackage{tikz}
\usepackage{adjustbox}
\usepackage{multirow}
\usepackage{pgfplots}
\usepackage{booktabs}
\usepackage[linesnumbered,ruled,vlined]{algorithm2e}
\usepackage{algpseudocode}
\usepackage[utf8]{inputenc}
\usepackage{tikz}
\usepackage{amssymb}% http://ctan.org/pkg/amssymb
\usepackage{pifont}% http://ctan.org/pkg/pifont
\usepackage{threeparttable}
\usepackage{tcolorbox}
\tcbuselibrary{breakable}
\usepackage{wasysym}
\usepackage{array}        % For better control of column formatting
\usepackage[table,xcdraw]{xcolor} % For row and column coloring
\usepackage{graphicx}
\usepackage{orcidlink}
\usepackage{booktabs,makecell,threeparttable,tabularx,array,wasysym}
\newcolumntype{L}[1]{>{\raggedright\arraybackslash}p{#1}}
\newcolumntype{C}[1]{>{\centering\arraybackslash}m{#1}}

\usepackage{amsthm}

\theoremstyle{definition}

\theoremstyle{remark}

\newcommand{\ie}{\textit{i.e., }}
\newcommand{\eg}{\textit{e.g., }}
\ifCLASSINFOpdf

\else

\fi

\begin{document}
	
	\title{Automated Byzantine-Resilient Clustered Decentralized Federated Learning for Battery Intelligence in Connected EVs}

	\author{

		\IEEEauthorblockN{Mouhamed Amine Bouchiha, ~\IEEEmembership{Member,~IEEE,} Abdelaziz Amara Korba,~\IEEEmembership{Member,~IEEE,}  Yacine Ghamri-Doudane,~\IEEEmembership{Senior Member,~IEEE} }

		\thanks{Mouhamed Amine Bouchiha is with SAMOVAR, Télécom SudParis, Paris, France; Abdelaziz Amara Korba is with the Department of Computer Science, German University of Technology in Oman (GUtech), Muscat, Oman. , and Yacine Ghamri-Doudane is with L3i, La Rochelle University, La Rochelle, France. E-mail: \{mbouchiha@telecom-sudparis.eu, abdelaziz.amarakorba@gutech.edu.om; yacine.ghamri@univ-lr.fr.\}}}

	% make the title area
	\maketitle

	\begin{abstract}

		Federated learning (FL) has emerged as a promising paradigm for managing electric vehicle (EV) battery data in connected and intelligent vehicles, enabling privacy-preserving tasks such as anomaly detection and capacity estimation. However, most existing frameworks rely on centralized aggregation schemes, which pose critical limitations in terms of security and trust. To address these challenges, we propose $ABC$-DFL, an automated Byzantine-resilient clustered decentralized federated learning (C-DFL) framework for connected EVs. The proposed incentive-driven C-DFL system replaces the central server with an open-permissioned blockchain, featuring a new dynamic Quorum Byzantine Fault Tolerance (QBFT) protocol and an oracle-based aggregation layer, to enhance trust, security, and automation. At the core of $ABC$-DFL lies FLECA (Filtered Layered Enhanced Clustering Aggregation), a robust hierarchical aggregation protocol that mitigates Byzantine attacks by having each EV filter malicious updates using an adaptive threshold based on deviations from its reference model update. \color{black} Oracle nodes, responsible for inter-group aggregation, employ robust clustering to isolate and aggregate model updates from trustworthy EV groups. Comprehensive experimental evaluations demonstrate that FLECA matches FedProx convergence under benign conditions and significantly outperforms existing defenses with attack impact scores below 0.10 in adaptive adversarial scenarios. Furthermore, several learning experiments with multitask models confirm the effectiveness and fairness of the incentive mechanism.  Finally, on-chain and off-chain benchmarks validate the practicality of $ABC$-DFL.
		
	\end{abstract}
	
	
	\IEEEpeerreviewmaketitle
	
	\section{Introduction}
	
	\IEEEPARstart{S}{mart} battery management plays a critical role in optimizing electric vehicle (EV) performance within intelligent transportation systems. State-of-the-art approaches leverage advanced data analytics and machine learning to improve battery diagnostics and capacity estimation, thereby enhancing safety and operational efficiency~\cite{abbaraju2024novel}. Accurate diagnostics and capacity estimation allow continuous monitoring of battery health, enabling the detection of faults such as overvoltage, overheating, and cell imbalance, while evaluating state-of-charge (SoC) and state-of-health (SoH) to maximize energy utilization and extend battery lifespan~\cite{iot2024battery, demirci2024review}. These capabilities are essential for robust integration of EVs into modern transportation infrastructure, supporting applications such as smart charging and vehicle-to-grid (V2G) systems~\cite{zhang2024survey}.  However, current EV battery intelligence approaches typically use centralized computing, sending raw data to cloud servers due to limited onboard processing power~\cite{li2021privacy}. This creates high latency and privacy risks (\ie driving behavior, usage patterns, and degradation profiles)~\cite{tvtflsoh, naresh2024privacy, han2024source}, despite enabling comprehensive pattern modeling.
	\color{black}
	%Current approaches for EV battery intelligence often rely on centralized computing frameworks, where data from EV Battery management systems (BMSs) is transmitted to cloud servers for processing due to the limited computational capabilities of onboard systems~\cite{li2021privacy}. However, this centralized approach introduces challenges, including high data transmission latency and privacy concerns. While centralized systems can model comprehensive battery data patterns, they are hindered by slow data collection and privacy risks, as EV users may be reluctant to share sensitive battery data~\cite{abbaraju2024novel}.
	
	Federated Learning (FL)~\cite{mcmahan2017communication} has been adopted as a semi-decentralized paradigm in Intelligent Transportation Systems (ITS)~\cite{li2021privacy, zhang2024survey} and, specifically for EV battery intelligence, to enable collaborative model training while preserving data privacy~\cite{tvtflsoh, smtfltce26}. However, traditional FL frameworks are vulnerable to single points of failure and lack transparency and verifiability in aggregation~\cite{srvfl2024, inferenceDSC2023}. To overcome these limitations, recent research has shifted towards Fully Decentralized Federated Learning (DFL). In DFL, EVs communicate directly in a peer-to-peer (P2P) manner, eliminating the central server and allowing EV to perform aggregation through direct model update exchange~\cite{wang2024peer, tvtbflev, zhang2024survey}. 
	
	Despite its advantages, DFL faces significant challenges, including susceptibility to poisoning attacks, where malicious participants corrupt local training data ~\cite{Virat2021, wang2021feature} or models~\cite{shejwalkar2021manipulating}. Existing Byzantine robust aggregations, designed primarily for server-assisted FL, are less effective in fully decentralized settings, where the absence of a central trusted aggregator complicates detection of malicious updates. Additionally, DFL suffers from high communication overhead and slower convergence due to inconsistent updates and varying battery data distributions across clients~\cite{srvdfl2024}. To address these limitations, we propose a novel \textit{Clustered} DFL (C-DFL) framework for battery intelligence in connected EVs. Our framework improves scalability, Byzantine resilience, and automation by introducing a trustless, hierarchical communication architecture. Our key contributions are:
	\begin{itemize}
		\item \textbf{$ABC$-DFL\footnote{\href{https://github.com/mohaminemed/ABC-DFL}{https://github.com/mohaminemed/ABC-DFL}}}: A Byzantine-resilient clustered decentralized federated multi-task learning framework built on a scalable \textit{open-permissioned blockchain}, complemented by an \textit{oracle network} and \textit{off-chain storage}. The blockchain enforces trust and security through integrated reward and reputation mechanisms, while oracles and off-chain storage nodes offload computation and storage to enhance scalability and automation beyond classical blockchain-based FL.
		
		\item \textbf{Dynamic Reputation-Based QBFT Protocol}: A novel blockchain protocol that introduces dynamic, reputation-weighted committee rotation and leader election via Verifiable Random Functions (VRFs). This directly addresses the static committee vulnerability and predictable leadership of standard QBFT, which are susceptible to targeted attacks, thereby ensuring fairness, liveness, and Byzantine robustness while preventing adversarial capture.
		
		\item \textbf{FLECA (Filtered Layered Enhanced Clustering Aggregation)}: A novel two-stage robust aggregation protocol that uniquely combines EV-side reference filtering with oracle-side robust clustering. FLECA fills the critical gap in decentralized settings by enabling Byzantine-resilient aggregation without any trusted aggregator, mitigating both isolated and colluding poisoning attacks.
		\color{black}
		
	\end{itemize}
	\color{black}
	
	The remainder of the paper is organized as follows. \S\ref{sec:relatedWork} reviews related work. \S\ref{sec:problemdef} defines the problem. \S\ref{sec:proposedframework} presents the proposed $ABC$-DFL framework. \S\ref{sec:dqbft} presents the integrated lightweight dynamic QBFT consensus protocol, while \S\ref{sec:fleca} describes the proposed FLECA protocol. \S\ref{sec:theoriticalanalysis} provides theoretical analysis, and \S\ref{sec:evaluation} reports the performance evaluation and experimental results. Finally, \S\ref{sec:conclusion} concludes the paper.
	
	\color{black}
	
	\section{Related Work} \label{sec:relatedWork}
	
	\color{black}
	
	%\subsection{Federated Learning for EV Battery Intelligence}

	Existing research on EV battery intelligence primarily relies on centralized machine learning for state-of-charge (SoC), state-of-health (SoH) estimation, and anomaly detection \cite{sultan2025enhancing,das2024analyzing,kumari2023baybfed}. While these approaches achieve high predictive accuracy, they inherently assume full data centralization, leading to privacy leakage risks, single points of failure, and significant communication overhead.
	
	Federated learning (FL) \cite{mcmahan2017communication} has been proposed to alleviate these concerns by keeping raw battery data local to EVs. Ensemble-based and personalized FL methods \cite{abbaraju2024novel,wang2024adaptive} improve learning performance under data heterogeneity but rely on a trusted central server for aggregation. Similarly, semi-decentralized and secure aggregation schemes \cite{wong2024decentralized,lopez2024rul,zhong2024lithium,han2024source} introduce encryption or adaptive mechanisms, yet still assume honest aggregators and do not consider Byzantine participants. Fully peer-to-peer FL \cite{wang2024peer} removes the central server but sacrifices scalability and provides only limited robustness guarantees. Crucially, across these works, adversarial behavior is either absent from the threat model or limited to passive privacy leakage. 
	\color{black}
	%\subsection{Decentralization and Trust in Connected EVs and ITS}
	
	Beyond battery analytics, FL has been applied to intelligent transportation systems (ITS) and vehicular edge computing (VEC) to enable privacy-preserving learning \cite{li2021privacy,liu2022mobile} with cooperative participants and trusted infrastructure. Furthermore, Blockchain has been explored as a trust anchor for FL in ITS \cite{liu2022blockchain,abdel2021federated}. Liu et al.~\cite{liu2022blockchain} integrate blockchain and digital twins to secure IoV communications, while Abdel-Basset et al.~\cite{abdel2021federated} propose miner-validated FL for intrusion detection. Although these systems enhance auditability and trust, they focus on centralized or single-layer aggregation and do not address scalability, hierarchical adversaries, or Byzantine behavior at multiple levels.
	
	In contrast to prior work, \textit{ABC}-DFL introduces a fundamentally different design point. Rather than relying on a trusted server or flat peer-to-peer topology, \textit{ABC}-DFL adopts a \emph{clustered fully decentralized federated learning architecture} for large-scale EV environments. The framework explicitly models and mitigates Byzantine behavior at both the intra (EV-level) and inter-cluster (group-level) aggregation stages.

	Distinctively, \textit{ABC}-DFL combines four key innovations that fill specific gaps in prior work: 
	(i) A \textit{C-DFL trustless verification} based on blockchain and oracles eliminates reliance on any trusted aggregator while solving the scalability limitation of pure on-chain aggregation~\cite{tvtbflaziz}.
	(ii) \textit{Dynamic QBFT consensus} with reputation-proportional committee rotation using VRFs, addressing the static committee vulnerability and predictable leadership of standard QBFT~\cite{cquorum}; 
	(iii) \textit{FLECA} provides a novel two-stage filtering protocol that couples EV-side reference checks with oracle-side clustering, solving the core decentralization challenge: enabling Byzantine detection when no trusted aggregator with global perspective exists; 
	(iv) \textit{Incentive-aware participation} employs a Gompertz-based reputation mechanism with asymmetric reward-penalty, mitigating free-riding and subtle manipulation unaddressed by averaging-based incentives.

	To the best of our knowledge, no existing FL framework for EV battery intelligence jointly addresses Byzantine robustness, full decentralization, and trustless aggregation under an explicit Byzantine threat model. Table~\ref{tab:comparison} summarizes this separation by contrasting threat assumptions, trust models, architectural choices, and scalability guarantees across prior art and \textit{ABC}-DFL.
	
	%Specifically, ABC-DFL makes the following technical contributions beyond prior art:
	
	%\begin{itemize}
	%\item \textbf{Dynamic QBFT for clustered FL:} we introduce a reputation-aware QBFT variant that dynamically reconfigures validator sets across FL rounds, enabling consensus despite malicious CSs—whereas prior blockchain-FL systems assume static or fully trusted validators.
	
	%\item \textbf{FLECA aggregation:} we propose a two-stage clustered robust aggregation that tolerates Byzantine EVs and Byzantine clusters simultaneously, filling the gap left by existing FL defenses that operate only at a single aggregation layer.
	
	%\item \textbf{Oracle-based trustless aggregation:} we design decentralized oracles that verify model updates and reputation scores on-chain, removing the trusted aggregator assumption present in all existing EV-FL frameworks.
	
	%\item \textbf{Incentive-compatible participation:} we couple robustness with an incentive mechanism that penalizes detected Byzantine behavior and rewards consistent contributors, whereas prior works treat security and incentives independently.
	%\end{itemize}
	
	\begin{table*}[t]
		
		\centering
		\caption{ Comparison of ABC-DFL with representative FL frameworks for EV battery intelligence.}
		\label{tab:comparison}
		\setlength{\tabcolsep}{3pt}
		\renewcommand{\arraystretch}{0.9}
		\begin{threeparttable}
			\begin{tabularx}{\textwidth}{L{0.17\textwidth} C{2.6cm} C{2.5cm} C{2.6cm} C{2.4cm} C{2.3cm}}
				\toprule
				\textbf{Reference} &
				\textbf{Byzantine Threats} &
				\textbf{Privacy Protection} &
				\textbf{Scalability} &
				\textbf{Aggregation Levels} &
				\textbf{Trust Model} \\
				\midrule
				
				Abbaraju et al.~\cite{abbaraju2024novel} &
				None &
				Local training &
				Centralized bottleneck &
				Single &
				Trusted server \\
				
				Han et al.~\cite{han2024source} &
				None (domain shift) &
				GMM abstraction &
				Centralized FTL &
				Single &
				Trusted server \\
				
				Wang et al.~\cite{wang2024peer} &
				None &
				Local training &
				Flat P2P DFL &
				Single &
				Peer trust \\
				
				Wang et al.~\cite{wang2024adaptive} &
				None  &
				Local training &
				Centralized bottleneck &
				Single &
				Trusted server \\
				
				Wong et al.~\cite{wong2024decentralized} &
				None &
				Local training &
				Semi-centralized FL &
				Single &
				Trusted coordinator \\
				
				Lopez et al.~\cite{lopez2024rul}&
				Model inversion only &
				HE scheme &
				HE overhead &
				Single &
				Trusted server \\
				
				Zhong et al.~\cite{zhong2024lithium} &
				None &
				Local training &
				Centralized FL &
				Single &
				Trusted aggregator \\

				\midrule
				\textbf{ABC-DFL (ours)} &
				\textbf{Byzantine EVs + Groups} &
				\textbf{Differential Privacy + Regularization} &
				\textbf{Hierarchical Clustered-DFL} &
				\textbf{Two-level (EV/Group)} &
				\textbf{Trustless (BC + Oracles)} \\
				
				\bottomrule
			\end{tabularx}
			
			\begin{tablenotes}[flushleft]\footnotesize
				\item Abbrev.: \textit{FEL = Federated Ensemble Learning; FTL = Federated Transfer Learning; GMM = Gaussian Mixture Model; HE = Homomorphic encryption.}
				\item \textit{Byzantine threats refer to malicious EVs and/or aggregators performing model poisoning or collusion.} %Prior works address privacy or domain shift under benign assumptions, whereas ABC-DFL jointly provides hierarchical scalability and Byzantine robustness with trustless aggregation.}
		\end{tablenotes}
	\end{threeparttable}
\end{table*}

\color{black}

\section{Problem Definition} \label{sec:problemdef}

This work addresses the challenge of designing a federated learning framework for EV battery intelligence that is simultaneously fully decentralized, scalable, and Byzantine-resilient.

The transition to fully decentralized federated learning (DFL), while essential for eliminating the single point of failure and trust dependency of a central server, introduces critical and interconnected research challenges that remain unaddressed in the existing literature.

Foremost among these is a fundamental mismatch between existing Byzantine defenses and the decentralized environment. Robust aggregation rules are designed for a central aggregator with a complete global view of all updates. In a fully decentralized peer-to-peer setting, no single entity possesses this comprehensive perspective. Applying these rules locally at each EV is not only computationally expensive but also less effective due to the limited local view. The absence of a trusted authority to execute filtering is the core of this problem.

Furthermore, there is an inherent tension between decentralization and practical scalability. A purely peer-to-peer DFL system~\cite{wang2024peer} generates communication overhead that becomes prohibitive for large-scale EV fleets. This naive approach is unsuitable for the envisioned scale of future intelligent transportation systems, necessitating a novel design that reduces communication load while preserving the benefits of decentralization. Consequently, the core research problem we address is: \textit{How can we design a fully decentralized federated learning framework for EV networks that provides verifiable Byzantine resilience without a trusted central entity, while ensuring practical scalability and fair incentives?}

This work proposes a Clustered DFL (C-DFL) architecture that organizes participants into hierarchical groups. While this design inherently improves scalability by localizing most communication within clusters, it also introduces new challenges: enabling Byzantine-robust aggregation within each group in the absence of a trusted leader, coordinating robust aggregation across groups without a central server, and guaranteeing auditability and accountability throughout the learning process.

\color{black}

\section{$ABC$-DFL Framework} \label{sec:proposedframework}

This section presents the $ABC$-DFL framework, which integrates decentralized learning with hierarchical Byzantine-resilient aggregation. The framework ensures robustness, scalability, and communication efficiency for battery data management across CSs where EVs are parked. Table~\ref{tab:notations_abcdfl} summarizes the key symbols and notations used throughout the paper.

\begin{table}[t]
	\centering
	\caption{ Summary of key symbols and notations.}
	\setlength{\tabcolsep}{3pt}
	\begin{tabular}{m{3.0cm} m{5.0cm}}
		\toprule
		\textbf{Notation} & \textbf{Description} \\
		\midrule
		$ABC$ & Automated Byzantine-robust Clustered  \\
		$ASC$ & Access smart contract \\
		$MSC$ & Model smart contract \\
		$CID$ & Content IDentifier \\
		$VRF$ & Verifiable Random Function \\
		$E$ & Per-task EVs count \\
		$k$ & Per-task EV group size \\        
		$\mathcal{V}$ & Dynamic QBFT committee size \\
		$IM_s^t$ & Intermediate model of group $s$ at round $t$ \\
		$GM^t$ & Global model at round $t$ \\
		$r_s$ & reward earned by group $s$ \\
		$R_s$ & $CS_s$ reputation score  \\
		$T$ & Total number of communication rounds \\
		$\mathcal{D}_{si}$ & Local dataset of EV $EV_i$ at $CS_s$ \\
		$\Delta w_{si}$ & Local model update from EV $EV_i$ at $CS_s$ \\
		$C_1, C_2, ..., C_p$ & $p$ clusters of updates formed by FLECA \\
		$\beta$ & Tolerance to statistical dispersion \\
		$\kappa$ & Temporal tightening factor \\
		$\lambda(t) = \frac{t}{T}$ & Monotonically increasing scheduling \\
		$m$ & Number of malicious EVs per group \\
		$\rho$ & Churn rate (fraction of EVs leaving/joining) \\
		\bottomrule
	\end{tabular}
	\label{tab:notations_abcdfl}
\end{table}

\subsection{System Model} 
\label{sec:systemmodel}

The $ABC$-DFL framework operates under a partially synchronous network model \cite{partialsynch} with five core entities:

\noindent\textbf{Model Publishers (MP).} initiate FL tasks (\ie SoC estimation) and publish them on-chain to recruit charging stations.

\noindent\textbf{Training Agents (TA).} EVs train local models $LM$ on private battery data and share updates within their groups.

\noindent\textbf{Intermediate Aggregators (IA).} CSs aggregate local models from EVs, producing intermediate models $IM$ validated through a group consensus before on-chain recording.

\noindent\textbf{Committee Members (CM).} (Validators) ensure transaction integrity using the dynamic QBFT protocol (\S\ref{sec:dqbft}), maintaining ledger security and consistency.

\noindent\textbf{Global Aggregators (GA).} (Oracles) perform robust global aggregation of $\{IM_s\}$ to produce a global model $GM$, and automate reward/reputation updates via off-chain evaluation.

Both intra- and inter-group communications operate with \textit{partial synchrony} \cite{partialsynch}. This means that message delays are bounded by $\Delta$ in the long term, to ensure safety and liveness.

\subsection{Threat Model} \label{sec:threatmodel}

We consider an \textit{adaptive, colluding, and computationally bounded} adversary that may corrupt EVs, MPs, TAs, IAs, and GAs at arbitrary training rounds. The adversary is aware of the aggregation, clustering, and incentive mechanisms and may attempt coordinated attacks across multiple groups.

As in standard Byzantine fault-tolerant consensus and robust federated learning, ABC-DFL operates under the \textit{minimal honest-majority} conditions required for correctness (\ie indistinguishability between honest and malicious updates). Specifically, within each CS group, \textit{fewer than half} of EVs are Byzantine, and globally \textit{fewer than half} of CS groups are malicious. Importantly, ABC-DFL does not assume perfect detection of malicious behavior within each group. When local \textit{honest-majority} conditions are violated, the framework degrades gracefully by isolating affected groups via inter-group clustering and incentive mechanisms, preventing global model corruption.

For the blockchain and oracle layers, we adopt the standard \textit{partial synchrony}~\cite{partialsynch} Byzantine fault-tolerant (BFT) assumption: the system can tolerate up to $f$ faulty nodes out of a total of $n$, requiring $n \ge 3f + 1$ to guarantee both safety and liveness~\cite{miller2016}. This assumption is \textbf{orthogonal} to the learning layer and is necessary to ensure consistency under partial synchrony. Under this model, $ABC$-DFL mitigates the spectrum of threats discussed in \S\ref{sec:sec_analysis}.

\color{black}

\begin{figure}[t]
	\centering
	\includegraphics[width=0.96\linewidth]{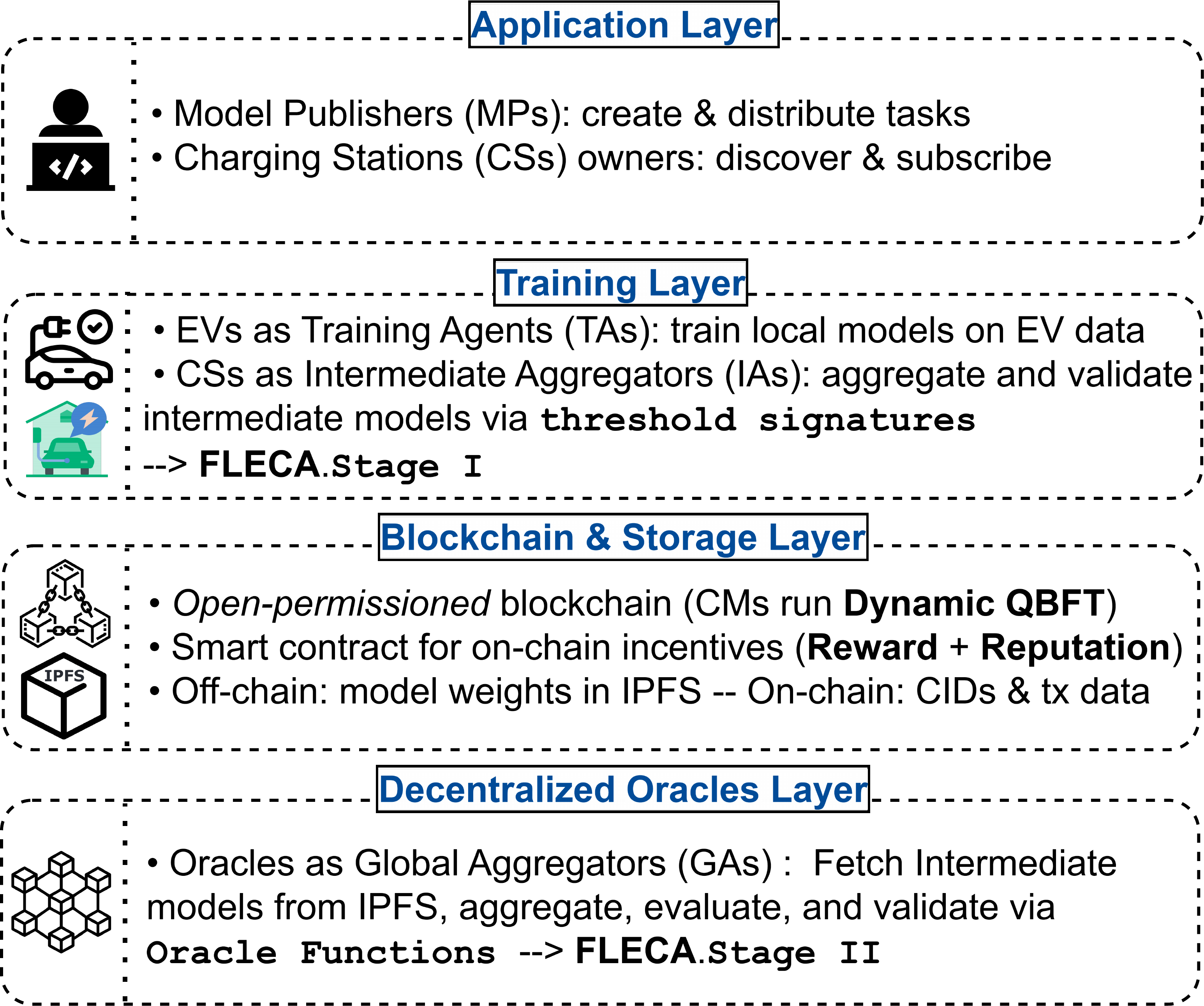} 
	\caption{An overview of $ABC$-DFL framework.}
	\label{fig:arch}
\end{figure}

\subsection{System Architecture}
\label{sec:archi}

The layered architecture of $ABC$-DFL, illustrated in Fig.~\ref{fig:arch}, orchestrates the interactions between the entities defined in §\ref{sec:systemmodel} across a secure and scalable workflow. 

\subsubsection{Application Layer}
Model Publishers (MPs) initiate the process by creating and distributing learning tasks (\ie anomaly detection, SoC estimation). CSs discover and subscribe to tasks matching their capabilities and interests.

\subsubsection{Federated Learning Layer}
Within each group $s$, EVs (a.k.a TAs) train local models on private data. Crucially, to ensure trustless collaboration within each group, EVs and CS run a threshold signature scheme~\cite{hints} to validate and agree upon a common intermediate model $IM_s$ before submission. 

\subsubsection{Blockchain \& Off-Chain Storage Layer}
An \textit{open-permissioned} blockchain~\cite{miller2016}, operated by a governing consortium (\ie a committee of transportation and renewable energy administrations), provides the trust anchor. Authorized CSs participate in a Dynamic QBFT consensus~\cite{cquorum} ($f < n/3$), ensuring instant finality and Byzantine tolerance. To maintain scalability, model weights are stored off-chain in IPFS; only their content identifiers (CIDs) and transaction data are recorded on-chain~\cite{autodfl}.

\subsubsection{Oracle Layer}
A Decentralized Oracle Network (DON)~\cite{don}, operated by the same consortium, bridges the blockchain and off-chain storage. Oracles (GAs) fetch intermediate models from IPFS, perform robust global aggregation, and evaluate contributions—offloading complex computations from the blockchain while preserving security.

\begin{figure}[t]
	\centering
	\includegraphics[width=1.0\linewidth]{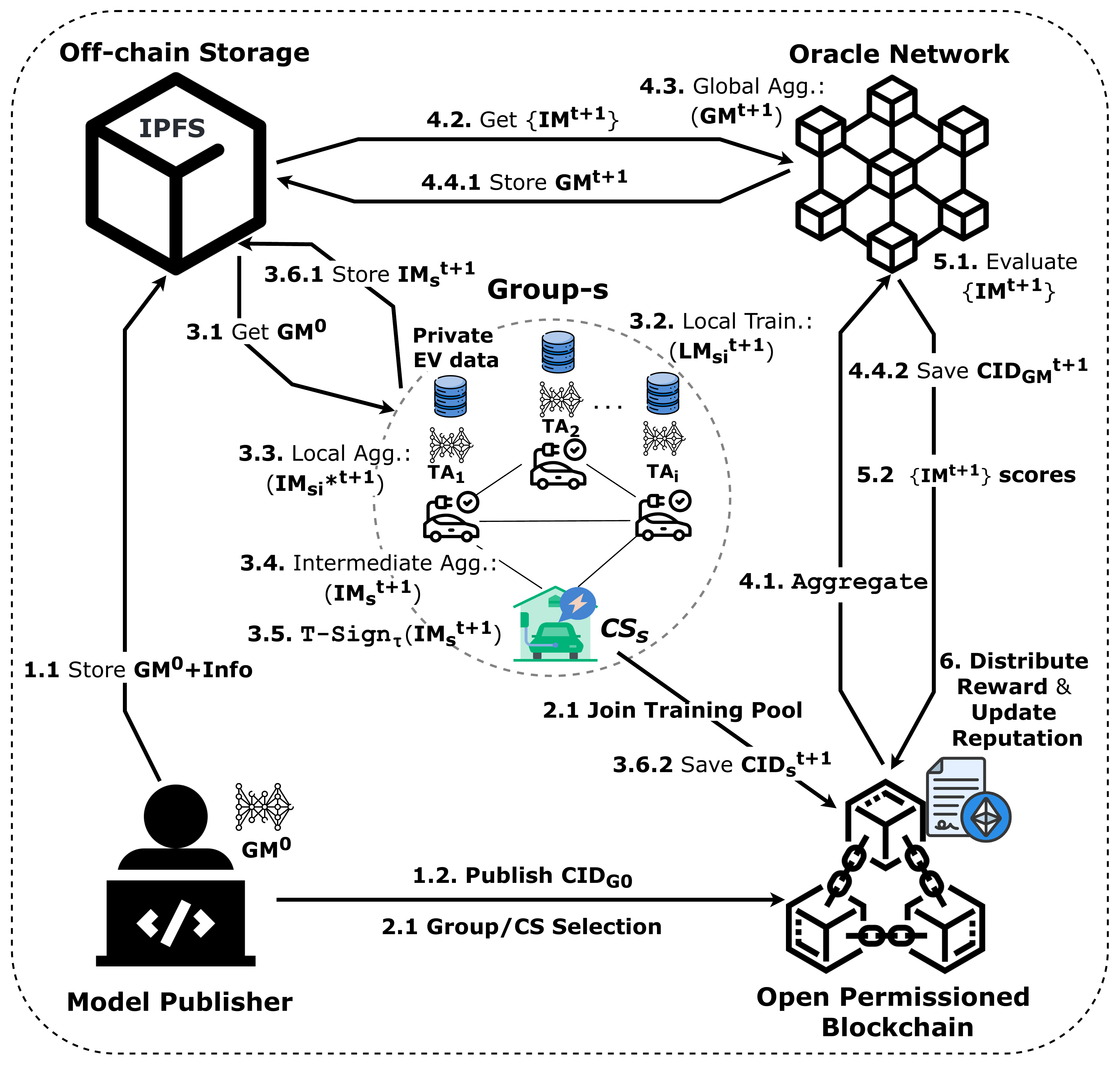} 
	\caption{$ABC$-DFL workflow: Comprises six phases: (1) model publishing by the MP via MSC, (2) CSs selection based on EVs data, deposits and reputation, (3–5) decentralized training through FLECA, where at each round $t$ EVs contribute to an intermediate model secured with threshold signing and differential privacy, followed by robust global aggregation via clustering with automated contribution evaluation at Oracles, and (6) on-chain reward distribution and reputation update.}
	\label{fig:workflow}
\end{figure}

\subsection{Design Workflow}

The workflow of $ABC$-DFL consists of the following phases, as depicted in Fig.\ref{fig:workflow}:

%\begin{enumerate}    
\subsubsection{Publish the Initial Model}(Steps 1.1-1.2)
An MP shares a single or multitask FL model, detailing the learning tasks, required number of CSs, and targeted performance / maximum number of rounds. The model details are stored off-chain on IPFS, and the resulting CID is recorded on the blockchain for transparency and reference.

\subsubsection{Select Charging Stations} (Steps 2.1-2.2) CSs/Groups can request to join a training pool of any shared model. Selection is executed on-chain either manually (by MP choice) or automatically (based on deposits and reputation scores~\cite{autodfl}). Reputation and deposits are managed by the model smart contract (MSC) to ensure transparency and resistance to sybil or malicious participation. 

\subsubsection{Train and Submit Intermediate Models} (Steps 3.1-3.5) Each selected CS retrieves the model architecture from IPFS and shares it with nearby EVs. EVs perform local training, exchange local models, and produce an intermediate model $IM_s$ (\S\ref{sec:fleca}). A threshold signature scheme ensures only jointly validated $IM_s$ values are accepted. Differential privacy~\cite{xiong2020comprehensive} with clipping is applied at the EV level, adding noise to weights ($\mathit{w}' = \mathit{w} + \epsilon$) to protect against data leakage (\S\ref{subsec:dp_analysis}). Validated IMs are uploaded to IPFS, and their CIDs are submitted on-chain via \texttt{submitIM}. The decentralized training and aggregation process, which ensures the secure recording of an IM, is detailed in Algo.\ref{algots_hl}.

\color{black}

\subsubsection{Global Aggregation} (Steps 4.1-4.4) Upon collecting submissions, the MSC triggers aggregation (\ie send \texttt{AGGREGATE} event). Intermediate models are filtered using robust clustering~\cite{Ester1996DBSCAN} to isolate outliers and poisoned updates. The largest cluster is then aggregated via FedAvg~\cite{mcmahan2017communication} to produce the global model ($GM$), which is stored on IPFS and referenced on-chain. Both intermediate and global aggregation are detailed in \S\ref{sec:fleca}.

\subsubsection{Evaluate Contributions} (Steps 5.1-5.2). Each IM is scored based on cluster membership. Models in the largest cluster receive positive updates, while divergent ones are penalized.

Let $C_1, C_2, ..., C_p$ denote the $p$ clusters obtained via the clustering algorithm, and let $|C_i|$ represent the number of models in cluster $C_i$. The largest cluster is identified as:

\[
C_{\text{max}} = \underset{i}{\arg\max} \; |C_i|
\]

Let $S^t_{s}$ be the current score of $CS_s$ that submitted $IM_s$. The score is updated based on $IM_s$’s cluster assignment, as follows:

\begin{equation} \label{equ:eval}
	S^{t+1}_{s} = 
	\begin{cases} 
		S^{t}_{s} + \Delta S_s & \text{if } IM_s \in C_{\text{max}} \\
		S^{t}_{s} - \Delta S_s & \text{if } IM_s \notin C_{\text{max}} \\
	\end{cases}
\end{equation}

Here, $\Delta S_s$ denotes the magnitude of the reward/penalty, which is statically defined and dynamically scaled according to the contribution of the model. Intermediate models outside of $C_{\text{max}}$ are considered divergent and receive a score penalty.

To capture both consensus strength and model quality, we define the score update of $IM_s$ as:

\begin{equation} \label{eq:score_update}
	\Delta S_s = \eta \cdot \frac{|C_i|}{\sum_{j=1}^p |C_j|} \cdot \exp\left(-\alpha \cdot d(IM_s, \mu_{C_{max}})\right),
\end{equation}

\noindent
where $|C_i|$ is the size of the cluster $C_i$ to which $IM_s$ belongs, $\mu_{C_{max}}$ is the centroid of the majority cluster, $d(IM_s, \mu_{C_{max}})$ is a distance metric (\ie Euclidean or cosine distance), $\eta$ is the maximum score update, $\alpha$ controls the sensitivity to distance-based penalization. Eq.\eqref{eq:score_update} ensures that model updates in large, tightly grouped clusters are rewarded more, while models in small or incoherent clusters are penalized based on both rarity and deviation.

\begin{algorithm}[t]
	
	\footnotesize
	\DontPrintSemicolon
	\caption{Secure Decentralized IM Sharing} \label{algots_hl}
	\KwData{Round $t$, Threshold $\tau$, Number of EVs $k$, modelID}
	\KwResult{On-chain TS verification and \textit{CID}($IM^{t+1}$) submission}
	
	\SetKwProg{Upon}{Upon}{ do}{end}
	
	\BlankLine
	
	% Step 1: Key Generation
	\Comment{At $t=0$: EVs collaboratively generate key shares}\;
	\Upon{receipt of \textsc{keyGenerationRequest}($k$, EVs)}{
		All $k$ EVs generate shared public key $PK$ and private key shares $\{SK_i\}_{i=1}^{k}$\;
		Agree on threshold $\tau$ for signature generation\;
	}

	\BlankLine
	% Step 2: Local Training & Aggregation
	\Comment{Each EV trains locally and aggregates intermediate models}\;
	\For{each $EV_i$}{
		Train local model $LM^{t+1}_{i}$ using local data\;
		Share $LM^{t+1}_{i}$ with neighbors\;
		Perform local aggregation to produce $IM^{*t+1}_{i}$\;
	}
	
	\BlankLine
	% Step 3: Consensus & Signing
	\Comment{EVs agree on global intermediate model $IM^{t+1}$}\;
	\For{each EV $i$}{
		Generate partial signature $sig_i$ on $IM^{t+1}$ \tcp*{Appendix~A : Alg.1}
	}
	\lIf{at least $\tau$ EVs provide partial signatures}{
		Aggregate partial signatures $\{sig_i\}_{i=1}^{\tau}$ to form $Sig$ \tcp*{Appendix~A : Alg.1}
	}
	
	\BlankLine
	% Step 4: Blockchain Submission
	\Comment{Submit the verified intermediate model on-chain}\;
	Call MSC.\texttt{submitIM} ($Sig$, $\textit{CID}(IM^{t+1})$, $modelID$, $t$) \tcp*{On-chain verification in Appendix~A: Alg.2}
	
\end{algorithm}

\subsubsection{Reward Distribution} (Step 6) Reward allocation and reputation updates are performed on-chain to ensure transparency and accountability. Rewards are allocated proportionally to CSs according to how often their IMs were within the majority cluster.

Let $r_i$ denote the reward assigned to $CS_s$, and $n_s$ represent the number of times the model $IM_s$  was selected in the largest cluster. The total reward for $s$  is proportional to $n_s$. The reward function is formulated as $r_s = b \cdot n_s$, where $b$ is a reward multiplier. CSs whose models appear more frequently in the largest cluster will receive higher rewards. The total reward for all CSs in a round is normalized to ensure the total rewards remain within a specific budget:

\begin{equation} \label{equ:reward}
	r_s^{\text{normalized}} = \frac{r_s}{\sum_{s=1}^{|\mathcal{S}|} r_s} \cdot Reward_{\text{total}}
\end{equation}

Where $Reward_{\text{total}}$ is the total reward budget (deposits locked by MP + CSs) and $|\mathcal{S}|$ is the number of CSs. Finally, CSs' reputation scores are updated using the model's final trust score $S_{s}$ (Eq.\ref{equ:eval}). We define the Gompertz transform $G(x) = \exp\big(- b \exp(- c x)\big)$. Then, the reputation update with forgetting factor $\alpha \in (0,1)$ is
\begin{equation}
	R_{s}^{\mathrm{new}} \;=\; (1-a)\,R_{s}^{\mathrm{old}} \;+\; a \, G(S_{s}).
	\label{eq:gompertz_update}
\end{equation}

Eq.\ref{equ:reward} and Eq.\ref{eq:gompertz_update} ensure fairness in reward distribution and reputation update (see \S\ref{subsec:incentive_analysis}). CSs with persistently low rewards or reputations are de-prioritized in future selections and may be subject to slashing. 
\color{black}
\section{Dynamic QBFT with Reputation-Based Committee and Leader Rotation}\label{sec:dqbft}

The blockchain layer of $ABC$-DFL requires a consensus protocol that is not only Byzantine fault-tolerant but also fair, efficient, and aligned with the framework's incentive model. The standard QBFT protocol~\cite{cquorum}, while robust, operates with a static validator set and deterministic leader rotation, making it vulnerable to targeted attacks and unable to leverage the reputation system built within $ABC$-DFL. This section introduces our dynamic reputation-based extension to QBFT, a core contribution that enhances the security, fairness, and decentralization of the entire system.

\subsection{Background: Standard QBFT Protocol}

The Quorum-based Byzantine Fault Tolerance (QBFT) protocol~\cite{cquorum} enables a fixed set of $n$ validators to reach consensus on a block chain. In each round, a designated \emph{leader} proposes a block. The other validators (replicas) then exchange \emph{prepare} and \emph{commit} messages. A block is finalized once a supermajority quorum of $2f+1$ out of $3f+1$ validators (where $f$ is the maximum number of Byzantine nodes) commits to it. Leader election typically follows a simple round-robin schedule among the static committee. While this design ensures simplicity and safety, it has critical limitations:
\begin{itemize}
	\item \textbf{Static Committee:} It cannot incorporate new reputable participants or remove misbehaving ones, leading to potential centralization and reducing Sybil resistance.
	\item \textbf{Predictable Leadership:} Deterministic leader rotation makes it easy for adversaries to anticipate and target the next leader, compromising liveness.
	\item \textbf{Decoupled from Application Incentives:} The consensus process is agnostic to the performance and reputation of participants within the federated learning tasks.
\end{itemize}
We address these limitations by introducing a dynamic reputation-proportional, verifiably random committee and leader rotation.

\subsection{Dynamic Randomized Committee Formation}
\label{ssec:comform}

At the beginning of each epoch (a fixed interval of $p$ rounds), a new committee is formed. This process, detailed in Algorithm~\ref{algo:committee}, ensures fairness, verifiability, and resistance to manipulation. It consists of the following steps:

\begin{enumerate}
	\item \textbf{Seed generation.} The leader derives a seed from the hash of the \textit{last finalized block} $\mathcal{B}^*$, ensuring resistance against manipulation of the randomness source.
	\item \textbf{Randomness via VRF.} Using a long-term key pair of a verifiable random function~\cite{vrf}, the leader computes a verifiable random value $r_1$ and proof $\pi$ with
	\[ (r_1, \pi) \leftarrow \mathsf{VRF}_{sk}(seed).
	\]
	Both $r_1$ and $\pi$ are published, enabling all participants to verify correctness of the VRF output. 
	\item \textbf{Randomness extension.} A sequence of $m-1$ additional random values is generated iteratively by hashing: $r_{i+1}=\mathsf{Hash}(r_i)$. This yields $\mathcal{R}=\{r_1,\dots,r_m\}$.
	\item \textbf{Reputation-proportional selection.} Each random value $r_i$ is normalized into a uniform variable $u_i \in [0,1]$, then mapped to one validator using the \textit{inverse cumulative distribution function (CDF)} over the reputation weights $\{rw_j\}$ of all candidates. Concretely, let cumulative weights be
	\[
	F_j=\frac{\sum_{t=1}^j rw_t}{R_W}, \quad R_W=\sum_{j=1}^n rw_j,
	\]
	where $n$ is the number of candidates. The selected validator is the unique index $j$ with $F_{j-1}\le u_i <F_j$.  
	\item \textbf{Committee output.} The set of $\mathcal{V}$ selected validators, together with proof $\pi$, defines the next committee.
\end{enumerate}

This design guarantees that committee selection is (i) \textit{fair}, being proportional to reputation, (ii) \textit{verifiable}, since the VRF proof is publicly checkable, and (iii) \textit{resistant to leader bias}, as the seed derives from a finalized block and cannot be manipulated by grinding.  

\begin{algorithm}[t]
	\footnotesize
	\DontPrintSemicolon
	\SetKwFunction{FVRF}{VRF}
	\SetKwFunction{FHash}{Hash}
	\SetKwFunction{FInverseCDF}{InverseCDF}
	\SetKwProg{Fn}{Function}{:}{}
	\caption{Dynamic Committee Selection \label{algo:committee}}
	\KwData{Reputation weights $\{rw_1,\dots,rw_n\}$ for $n$ candidates, committee size $\mathcal{V}$, last finalized block $\mathcal{B}^*$.}
	\KwResult{Selection of  $\mathcal{V}$ committee members with proof $\pi$.}
	\Begin{
		Compute seed from last finalized block: $seed \leftarrow \FHash(\mathcal{B}^*)$; \\
		Compute VRF output and proof: $(r_1, \pi) \leftarrow \FVRF_{sk}(seed)$; \\
		Initialize $\mathcal{R} \leftarrow \{r_1\}$; \\
		\For{$i \gets 2$ \KwTo  $\mathcal{V}$}{
			$r_i \leftarrow \FHash(r_{i-1})$; \\
			Add $r_i$ to $\mathcal{R}$; \\
		}
		Compute total reputation $R_W=\sum_{j=1}^n rw_j$; \\
		Construct cumulative distribution $F_j=\frac{\sum_{t=1}^j rw_t}{R_W}$ for $j=1,\dots,n$; \\
		Initialize committee $C \leftarrow \emptyset$; \\
		\ForEach{$r_i \in \mathcal{R}$}{
			Normalize: $u_i \leftarrow r_i / 2^b$ \tcp*{$b$ = bit length of VRF output}
			Select validator $v$ with smallest index $j$ such that $F_j > u_i$; \\
			Add $v$ to $C$; \\
		}
		\textbf{return} $(C,\pi)$; \\
	}
\end{algorithm}

\subsection{Dynamic Randomized Leader Election}

Once the committee $C$ is formed, a leader is elected for each round. Instead of deterministic round-robin, our protocol employs a reputation-weighted random election to further strengthen Sybil resistance and to reward trustworthy CSs with higher leader frequency.

For round $t$, the randomness $r_t$ (derived either from the epoch VRF output or the round hash) is mapped to the cumulative distribution of committee reputation scores. Specifically, let each committee member $c_j$ hold a reputation weight $rw_j$, and let $R_W=\sum_{j=1}^m rw_j$. A leader is selected by computing:
\[
Idx = r_t \bmod R_W,
\]
and choosing the first node $c_j$ such that
\[
\sum_{u=1}^j rw_u > Idx.
\]

\noindent This mechanism balances fairness and efficiency: honest and well-performing CSs are incentivized with higher leader probability, while randomness prevents leader predictability and reduces the risk of targeted attacks.

\section{Filtered Layered Enhanced Clustering Aggregation} \label{sec:fleca}

This section presents the proposed \textit{Filtered Layered Enhanced Clustering Aggregation} (FLECA) protocol, a Byzantine-resilient aggregation mechanism designed for our C-DFL system. The protocol employs \emph{two-stage filtering strategy} to mitigate poisoning threats: the first stage operates at \emph{EV level} (detecting and excluding malicious EVs), while the second is executed by \emph{Oracles} (detecting and filtering adversarially manipulated groups). This layered design enhances robustness against both isolated and colluding adversaries.

\subsection{Protocol Setup}

The FLECA protocol engages four $ABC$-DFL entities (\S\ref{sec:proposedframework})—TAs, IAs, CMs, and GAs—operating across three peer-to-peer overlays: (i) the \emph{group network}, (ii) the \emph{oracle committee}, and (iii) the \emph{open-permissioned blockchain}. The interactions between these entities provide a secure and verifiable model training with robust aggregation. The associated optimization problem is defined as:
\begin{equation}
	w^* = \arg \min_{w \in \Theta} F(w) = \frac{1}{|D|} \sum_{\zeta \in D} f(w, \zeta),
\end{equation}
where $\Theta \subset \mathbb{R}^d$ denotes the model parameter space of dimension $d$. The population risk $F(w)$ is approximated from distributed datasets $D$, and $f(w, \zeta)$ denotes the empirical loss on a sample $\zeta$ under model $w$.

\subsection{Two-Stage Filtering and Aggregation}

Poisoning adversaries may manipulate model updates to bias the global model. At the \emph{first stage}, EV-level filtering discards locally inconsistent models. At the \emph{second stage}, Oracles detect and exclude poisoned group-level updates. This layered defense is designed to counter both independent and colluding adversaries.

\subsubsection{Stage I: EV-Level Filtering and Aggregation}

Consider a $CS_s$ coordinating EVs $\mathcal{E}_s = \{EV_{s1}, \dots, EV_{sk}\}$. Each new round $t$, each EV $EV_{si} \in \mathcal{E}_s$ trains a local model:
\begin{equation}
	LM^{t+1}_{si} =  w_{si}^{t+1} = \arg \min_{w \in \Theta} f(w, \zeta_{si}), \quad \zeta_{si} \in D_{si},
\end{equation}
where $D_{si}$ is the dataset of EV $i$. Next, each EV $EV_{si}$ shares its model updates $\Delta w_{si}^{t+1} = w_{si}^{t+1} - w^{t}$ with its neighbors. 

Each EV performs adaptive similarity filtering by comparing received neighboring updates against its own reference update using a relative $\ell_2$-norm distance computed on task-specific output layers. For each neighboring update $\Delta w^{t+1}_{sk}$, the EV computes
\begin{equation}
	\label{eq:adaptive_score}
	s_k^t = \max_{\ell \in \mathcal{L}}
	\frac{\left\| \Delta w^{t+1}_{sk,\ell} - \Delta w^{t+1}_{si,\ell} \right\|_2}
	{\left\| \Delta w^{t+1}_{si,\ell} \right\|_2 + \varepsilon},
\end{equation}
where $\Delta  w^{t+1}_{si}$ denotes the EV’s reference update, $\mathcal{L}$ is the set of monitored output layers (classification and regression heads), and $\varepsilon$ is a small constant for numerical stability.

To robustly distinguish benign from malicious updates under data heterogeneity, the EV derives an adaptive acceptance threshold based on the median absolute deviation (MAD) of the similarity scores:
\begin{equation}
	\label{eq:adaptive_threshold}
	\theta^t =
	\frac{\mathrm{median}(s^t) + \beta \cdot \mathrm{MAD}(s^t)}
	{1 + \kappa \cdot \lambda(t)},
\end{equation}
where $\beta$ controls tolerance to statistical dispersion, $\kappa$ regulates temporal tightening, $\lambda(t)=\frac{t}{T}$ is a monotonically increasing scheduling function with $t$ the current communication round. A neighboring update is accepted if $s_k^t \leq \theta^t$. If no update satisfies this condition, the EV falls back to its own reference model. Accepted updates are then aggregated using FedAvg.
\color{black}
\begin{equation}
	IM^{*{t+1}}_{si} = \Delta w^{*t+1}_s = \frac{1}{|S_i^{t+1}| + 1} \Big(\Delta w_{si}^{t+1} + \sum_{j \in S_i^{t+1}} \Delta w_{sj}^{t+1} \Big),
\end{equation}
with $S_i^{t+1}$ the accepted set.  

\begin{algorithm}[t]
	%\scriptsize
	
	\caption{Majority voting} \label{algo:majvoting}
	\begin{algorithmic}[1] 
		\Require \(\{\Delta w_i, ID_i\}_{i\in\mathcal{E}_s}\) EV updates at station \(s\).
		\Ensure Final aggregated model $w^t$
		
		\State Count the occurrences of each model ID across all EVs.
		\State Identify strict majority IDs: \( \text{count} > |\mathcal{E}_s|/2 \)\;
		\State \lIf{no strict majority exists}{
			Select IDs with the maximum occurrence as a fallback.
		}
		\State Retrieve the models corresponding to the selected IDs.
		\State Aggregate the selected models using averaging.
		\State \Return aggregated intermediate model \(IM^{t+1}_s\)
		
	\end{algorithmic}
\end{algorithm}

Two variants are proposed for deriving the group-level update:

\noindent \textbf{Variant V1 (Majority Voting).}
\begin{equation}
	IM^{t+1}_s = \Delta w^{t+1}_s =  \text{MajorityVote}\{IM^{*{t+1}}_{si}; i \in \mathcal{E}_s\}.
\end{equation}
In this variant, each EV contributes a set of accepted model IDs. The majority voting procedure counts the occurrences of each model ID across EVs. If a strict majority exists (i.e., an ID appears in more than half of the EVs), it is selected. If no strict majority exists, a robust fallback selects the ID(s) with the maximum occurrence. The selected models are then aggregated via averaging to produce the station’s intermediate model. This ensures that aggregation is driven by the most widely supported updates.

\noindent \textbf{Variant V2 (Robust Clustering Filtering).}  
\begin{equation}
	IM^{t+1}_s = \Delta w^{t+1}_s = \text{ClusFedAvg}\{IM^{*{t+1}}_{si}; i \in \mathcal{E}_s\}.
\end{equation}

Variant V2 is specifically designed for deployment scenarios in which local EV configurations of the adaptive filtering mechanism
(\ref{eq:adaptive_threshold}) are \emph{heterogeneous}. To address this, each EV’s update is flattened into a task-specific vector representation. A density-based clustering method (HDBSCAN) is applied using a precomputed distance matrix between EV updates. The largest non-noise cluster, representing the most mutually consistent updates, is selected, and its members are averaged to obtain the station’s intermediate model. This approach mitigates the influence of outliers and adversarial updates while preserving the dominant patterns in the EV updates.

In both cases, $CS_s$ finalizes the intermediate model via a threshold signing scheme (Algo.~\ref{algots_hl}).

\subsubsection{Stage II: Oracle-Level Filtering and Aggregation}

Oracles retrieve intermediate models $\{w_s^{t+1}\}$ from IPFS and perform clustering-based outlier filtering. Poisoned CS-level models are often coherent but deviate collectively from benign updates. Detecting these requires density-based clustering methods robust to arbitrary shapes and adversarial noise.

The filtering algorithm uses robust \emph{HDBSCAN} for adaptive filtering under heterogeneous conditions. Let $\mathcal{S}$ denotes the set of CSs, the final global update is then:
\begin{equation}
	GM^{t+1} = \text{ClusFedAvg}\{\Delta w_s^{t+1}; s \in \mathcal{S}\},
\end{equation}

\begin{algorithm}[t]
	%\scriptsize
	
	\caption{Robust Clustering-based Filtering} \label{algo:clustering}
	\begin{algorithmic}[1] 
		\Require Model updates $\{\Delta w^{t}_1, \dots, \Delta w^t_n\}$
		\Ensure Final aggregated model $w^t$
		\State \lIf{\(|\mathcal{E}_s| \le 1\)}{ 
			\Return all updates
		}
		\State Flatten task-specific weights of each EV into vector \(V_i\)
		\State Compute pairwise distance matrix \(D\) between \(\{V_i\}\)
		\State Apply HDBSCAN clustering on \(D\)
		\State Identify the largest non-noise cluster \(C_{\max}\)
		\State Compute clustered FedAvg: 
		\(\bar{V} = \frac{1}{|C_{\max}|} \sum_{V_i \in C_{\max}} V_i\)
		\State Convert \(\bar{V}\) back to parameter form \(\Delta w^t\)
		\State \Return aggregated intermediate model update \(\Delta w^t\)
		
	\end{algorithmic}
\end{algorithm}

Unlike DBSCAN, which requires fixed parameters ($\epsilon$, $minPts$) and is hard to tune under heterogeneous data, HDBSCAN builds a density hierarchy, removing the need for a global $\epsilon$ and providing greater robustness to adversarially skewed updates, making it better suited for dynamic, non-stationary environments. We discuss HDBSCAN sensitivity to $minPts$ in \S\ref{sec:ablation_study}.
\color{black}
\section{Theoretical Analysis} \label{sec:theoriticalanalysis}
\quad This section analyzes $ABC$-DFL's complexity and $FLECA$'s convergence and security guarantees.

\subsection{Complexity Analysis}
We report per-round costs (computation and communication). Each round in C-DFL with FLECA proceeds in the following pipeline: (A) local EV–EV exchange and filtering, (B) EV $\rightarrow$ CS upload of accepted local updates, (C) CS-level aggregation (majority or clustering), (D) CS $\rightarrow$ Oracles/IPFS commit and oracle-level clustering, and (E) global commit.

\noindent Let \(\mathcal{S}\) be the number of CSs, \(\mathcal{E}_s\) the number of EVs served by station \(s\), and \(E=\sum_s \mathcal{E}_s\) the total number of EVs. Let \(d\) be the model parameter dimension, \(k\) the average neighbor-degree used in EV-to-EV exchanges (gossip radius), and \(O\) the number of oracle nodes. We measure communication in \emph{messages} and \emph{bytes} (model size is \(O(d)\) floats); computation is measured in elementary vector operations (cost $\propto$  to \(d\)).

\subsubsection{Communication across DFL systems}  
In fully blockchain-based FL~\cite{bellachia2025verifbfl}, each round requires $\Theta(E^2)$ messages and $\Theta(E^2 d)$ bytes. Graph-based P2P DFL~\cite{guo2021byzantine} reduces message complexity to $\Theta(E k)$ and byte complexity to $\Theta(E k d)$ per iteration, but repeated neighbor averaging increases bandwidth consumption and slows convergence due to the lack of global aggregation. In contrast, our Clustered-DFL organizes clients into clusters of size $k$ and applies hierarchical compression, which bounds long-range communication to $\Theta(\mathcal{S} d)$, achieving \textit{linear scaling with cluster size} while maintaining efficiency and convergence.

\subsubsection{Computation across aggregation methods}  
At the EV level, centralized filtering at a trusted CS incurs $O(\mathcal{E}_s^2 d)$ for M-Krum~\cite{blanchard2017machine}, $O(\mathcal{E}_s d)$ for Trimmed-Mean~\cite{yin2018byzantine}, and $O(\mathcal{E}_s^2 d)$ for FLAME~\cite{Flame} with HDBSCAN clustering and norm clipping/noise injection. Distributed filtering at EVs has $O(k^2 d)$ for local Multi-Krum and $O(k d \log k)$ for distributed Trimmed-Mean. UBAR~\cite{guo2021byzantine} applies a two-stage local filter (distance $O(k d)$ + loss pruning $O(k)$) within EV groups and adds $O(k d)$ inference per EV, assuming knowledge of the malicious fraction. FLECA further improves efficiency with $O(k d)$ per EV and CS-level majority voting or HDBSCAN clustering averaging $O(\mathcal{E}_s \log \mathcal{E}_s)$. For global aggregation, the complexity of each method is similar, with $\mathcal{E}_s$ replaced by $\mathcal{S}$; e.g., FLECA $O(\mathcal{S} \log \mathcal{S})$, M-Krum $O(\mathcal{S}^2 d)$.

\color{black}

\subsection{FLECA Convergence Guarantees}

We provide a non-convex convergence guarantee for FLECA that accounts for hierarchical robust aggregation (\S\ref{sec:fleca}) and a bounded Byzantine fraction (\S\ref{sec:threatmodel}).

\noindent\textbf{Assumptions.}
We adopt the following standard conditions.

\begin{enumerate}
	\item Each local loss \(f_i(\cdot)\) is \(L\)-smooth.
	\item Honest stochastic gradients have bounded variance:
	\(\mathbb{E}\|g_i(x)-\nabla F(x)\|^2 \le \sigma^2\).
	\item All client updates are clipped to norm \(C\).
	\item At both EV and CS aggregation layers, strictly fewer than half of contributors are Byzantine.
\end{enumerate}

Under Assumption~4, the hierarchical filtering mechanism of FLECA guarantees bounded aggregation bias. Specifically, Proposition~1 in \textbf{Appendix~C} shows that the global update at round \(t\) admits the decomposition
\[
w^t = \nabla F(x_t) + \xi_t + b_t,
\]
where \(\mathbb{E}[\xi_t\mid x_t]=0\), \(\mathbb{E}\|\xi_t\|^2\le\sigma^2\), and
\[
\|b_t\| \le B := r + \phi C,
\]
with \(\phi<1/2\) the Byzantine fraction and \(r\) the maximal dispersion of honest updates~\cite{blanchard2017machine, yin2018byzantine}.

\noindent\textbf{Proof sketch.}
The global model evolves as
\(x_{t+1}=x_t-\eta w^t\).
Using \(L\)-smoothness of \(F\), a standard descent lemma gives
\[
F(x_{t+1}) \le F(x_t)
-\eta\langle\nabla F(x_t), w^t\rangle
+\tfrac{L\eta^2}{2}\|w^t\|^2.
\]

Substituting the above decomposition of \(w^t\), taking conditional expectation, and bounding cross terms via Young’s inequality yields a descent inequality consisting of three components: a negative term proportional to \(\|\nabla F(x_t)\|^2\), a stochastic variance term proportional to \(L\eta\sigma^2\), and an additive term proportional to the squared aggregation bias \(B^2\).

Choosing a constant step-size \(\eta\le 1/(4L)\) ensures that the descent term dominates higher-order smoothness contributions. Summing over \(t=0,\dots,T-1\) and telescoping objective values gives
\[
\frac{1}{T}\sum_{t=0}^{T-1}\mathbb{E}\|\nabla F(x_t)\|^2
\le
\frac{2(F(x_0)-F^\star)}{\eta T}
+L\eta\sigma^2
+\tfrac{5}{2}B^2.
\]

The complete derivation and explicit constants provided in \textbf{Appendix~C} show that FLECA converges to a neighborhood of stationary points whose radius scales as \(O(B^2)\). Since \(B=r+\phi C\), the asymptotic error is explicitly controlled by the Byzantine fraction and clipping threshold. When honest updates are concentrated and \(\phi\) is small, FLECA recovers the standard stochastic convergence behavior while providing strong Byzantine resilience via hierarchical filtering.

\color{black} 
\subsection{System-wide Security Analysis} \label{sec:sec_analysis} Below, we discuss security threats and the corresponding defenses present in $ABC$-DFL.

\textbf{Sybil attacks.} Our system mitigates this by enforcing strict identity verification, with designated admins (\ie consortium members) exclusively authorized to add or remove users under consortium oversight. This is achieved through on-chain governance with a majority voting scheme, while the open-permissioned design enforces one update per CS per round (\texttt{submitIM}) via on-chain access control and per-task registration (\texttt{joinCSModel}). Consequently, Sybil-based amplification is effectively prevented unless an attacker compromises the consortium or the underlying blockchain layer (see \textbf{Appendix~A}).
\color{black}

\textbf{Poisoning and model manipulation.} FLECA provides \emph{layered} defense against coordinated poisoning:

\begin{enumerate}
	\item \textit{EV-level adaptive filtering.}
	Each EV rejects incoming models whose distance from its local model exceeds an adaptive threshold (Eq.~\ref{eq:adaptive_threshold}). The monotonically decreasing schedule forces adversaries to craft updates increasingly close to benign gradients, limiting effective perturbation.
	
	\item \textit{CS-level robust aggregation.}
	At each CS, EV updates are filtered via either majority-based connected components (v1) or clustering-based outlier removal (v2). Under the standard honest-majority condition (\(<50\%\) Byzantine EVs), this selects a benign-dominated subset, yielding bounded bias (Lemma~C.1 and Proposition~C.1).
	
	\item \textit{Inter-group clustering.}
	Oracles apply an additional clustering step across CS-level aggregates, isolating coordinated malicious groups. When fewer than half of CSs are Byzantine, hierarchical robustness composes, ensuring that the global update deviates from the true gradient by at most a bounded bias term \(B=r+\phi C\) (Theorem~C.1).
	
	\item \textit{Threshold signatures and content addressing.}
	EVs individually sign local updates; a CS cannot produce a valid intermediate model unless at least \(\tau\) EV signatures are collected. Aggregated models are stored off-chain (IPFS) and referenced on-chain via content hashes. Oracles verify both signature sets and hashes prior to acceptance, preventing model forgery, replay, or substitution.
\end{enumerate}
\color{black}
Together, these mechanisms guarantee that poisoning attacks cannot arbitrarily bias the global model unless honest-majority assumptions are violated. As shown in Theorem~C.1, the resulting learning dynamics reduce to biased stochastic gradient descent with explicitly bounded bias, yielding convergence to a neighborhood of stationary points whose radius scales as \(O((r+\phi C)^2)\).

\textbf{Free-riding and incentive manipulation.} Rewards are distributed at the end of each task according to the intermediate models' cluster membership (Eq.\ref{equ:reward}). Reputations are updated with a Gompertz transform $G(x) = \exp(-b \exp(-c S_{s}))$ and forgetting factor $a \in (0,1)$ (Eq.\ref{eq:gompertz_update})
where $S_{s}$ is a moving weighted score (Eq.\ref{equ:eval}) that combines cluster membership with distance-based penalization; and $b,c>0$ control the shape of the Gompertz curve. Reputation update ensures that models in large, coherent clusters are rewarded, while those in small inconsistent clusters are penalized. CSs with low rewards and reputations are de-prioritized in future tasks or may be subject to slashing, effectively mitigating free-riding.

\label{subsec:dp_analysis}
\textbf{Regularization and privacy leakage.} Benign EVs apply a weak Gaussian differential-privacy (DP) mechanism on clipped updates. Each EV clips its local update to \( \| \Delta w \|_2 \le C\) and releases
\(\tilde{\Delta w} = \Delta w + \mathcal{N}(0,\sigma^2 I)\).
Using the standard Gaussian mechanism, for target \((\varepsilon,\delta)\)-DP the noise scale must satisfy
\begin{equation} \label{eq:std_dp}
	\sigma \ge \frac{C\sqrt{2\log(1.25/\delta)}}{\varepsilon}.
\end{equation}
This provides composable privacy guarantees; in practice one chooses \(C,\varepsilon,\delta\) to trade utility and privacy (see \S\ref{sec:ablation_study}). Importantly, DP is applied primarily as a client-side regularization mechanism. This ensures that the added noise does not interfere with outlier detection.

\subsection{Incentive Compatibility and Strategic Robustness} \label{subsec:incentive_analysis}

We model the reward--reputation mechanism as a repeated game among a finite set of CSs (EV groups). Each CS chooses at every round whether to submit an \emph{honest} intermediate model (IM), incurring a higher
computation cost, or a \emph{deviating} IM (\ie free-riding or manipulated), incurring lower cost.

Rewards are allocated proportionally to the frequency with which a CS's IM appears in the majority cluster (Eq.~\ref{equ:reward}), while reputation is updated via a Gompertz-based transform with temporal smoothing (Eq.~\ref{eq:gompertz_update}). Reputation directly influences future selection probability and expected rewards.

Under standard assumptions on robust clustering (\ie \textit{honest IMs form the majority cluster with high probability if the fraction of deviating CSs is below a threshold}), honest behavior strictly dominates deviation in expected long-term utility. In particular, the expected gain from reduced computation costs under deviation is outweighed by (i) lower probability of reward allocation, (ii) accelerated reputation decay, and (iii) economic slashing. To discourage strategic manipulation (\eg reputation farming or slow free-riding), we further employ an \emph{asymmetric reputation update}, in which reputation increases slowly following honest behavior but decays rapidly upon low trust scores. A formal game-theoretic analysis and proof sketch establishing honest participation as a Nash equilibrium in the repeated game are provided in \textbf{Appendix~D}.
\color{black}

\section{Evaluation and Results} \label{sec:evaluation} 

This section presents a comprehensive evaluation of the $ABC$-DFL framework, assessing its performance across multiple dimensions including learning efficiency, robustness against poisoning attacks, incentive effectiveness, and overall scalability. We implemented a proof-of-concept of the system, with the source code for the FLECA protocol, smart contracts, and benchmarking tools available on \textbf{GitHub}\footnote{\url{https://github.com/mohaminemed/ABC-DFL}}. 
\color{black}
\subsection{Experimental Setup}

\subsubsection{Datasets and models} FLECA robustness is evaluated using the  EV battery diagnostics ``EVBattery'' dataset~\cite{he2022evbattery} , which addresses two critical operational needs: health anomaly detection to prevent catastrophic failures (\eg fire), and capacity estimation for accurate range prediction and battery management. This comprehensive dataset includes time-series data collected during fixed-duration charging sessions, along with associated meta-information for each recording. Each time-series snippet consists of 128 time-steps, capturing the detailed dynamics of the battery’s behavior throughout the charging process. The dataset records eight key features, including voltage, temperature, state-of-charge (SoC), and timestamps. In addition to the time-series data, the dataset provides meta-information such as a unique vehicle identifier, mileage at the time of data collection, snippet index, and labels for battery health and capacity. The health label is binary, with a value of \(1\) indicating anomalies (\ie lithium plating or thermal runaway) and \(0\) representing normal operation. The capacity label, expressed in ampere-hours (Ah), indicates the remaining battery capacity, ranging from 28.28 Ah to 46.23 Ah.  To simulate realistic non-IID distributions across EVs, the data is partitioned per EV using a Dirichlet distribution with concentration parameter \(\alpha\), and all partitions are generated using fixed global random seeds (42, 70, 84) to ensure reproducibility. This setup allows controlled heterogeneity in both feature distributions and label proportions, reflecting operational variability encountered in real-world EV deployments.
\color{black}

We evaluated several multitask models for joint anomaly detection (binary classification) and capacity estimation (regression):
\begin{itemize}
	\item \textit{MultiTaskLSTM:} LSTM encoder with dual output heads.
	\item \textit{MultiTaskbiLSTM:} Bidirectional LSTM with classification and regression heads.
	\item \textit{MultiTaskCNN:} Stacked 1D CNN layers with dual heads.
	\item \textit{MultiTaskGRU:} GRU-based encoder with dual heads.
\end{itemize}

\subsubsection{Simulated attacks} We examine multiple poisoning attacks, consisting of data poisoning attacks (Label Flipping~\cite{Virat2021}, Feature~\cite{wang2021feature}) and model poisoning attacks: (Gauss~\cite{blanchard2017machine}, Krum, Trim~\cite{fang2020local}), Adaptive~\cite{shejwalkar2021manipulating}, and Backdoor attacks (Badnets~\cite{BadNets}, Scaling~\cite{ModelReplacement}, Neurotoxin~\cite{neurotoxin}). The details of each attack are presented in \textbf{Appendix~E.1}.
\color{black}

\subsubsection{Evaluation metrics} We assess battery health anomaly detection using: %accuracy, precision, recall, F1-score, Max.TER, and AUROC: 

\begin{itemize}
	\item[] $\text{Accuracy} = \frac{TP + TN}{TP + TN + FP + FN}$\text{; Precision} $= \frac{TP}{TP + FP} $\text{;} \:\: \text{Recall} $ = \frac{TP}{TP + FN}$ $\text{; F1-Score} = 2 \times \frac{\text{Precision} \times \text{Recall}}{\text{Precision} + \text{Recall}}$
	\vspace{4pt}
	%$\text{; and TER} = \frac{FP + FN}{TP + TN + FP + FN}$
\end{itemize}

%We compute Max.TER by measuring the testing error rate (TER) of the final local model for each benign EV, and use the maximum error rate among all benign EVs to assess the robustness of FLECA.

For the battery capacity estimation task, we use the mean absolute error (MAE), the mean squared error (MSE), and the root mean squared error (RMSE):  

\begin{itemize}
	\item[]   \text{MAE} $= \frac{1}{N} \sum_{i=1}^{N} | y_i - \hat{y}_i | $ \text{; MSE} $= \frac{1}{N} \sum_{i=1}^{N} (y_i - \hat{y}_i)^2$\text{; and} \:\:  \text{RMSE} $= \sqrt{\text{MSE}}$
\end{itemize}

We define the Attack Impact Score (AIS) to measure of the impact of a poisoning attack on both anomaly detection and capacity prediction. It consists of:

1. Anomaly degradation: Measures the reduction in anomaly detection performance (\eg Accuracy, F1-Score):
\begin{equation}
	AIS_{\text{A}} = \frac{\text{anomaly}_{\text{benign}} - \text{anomaly}_{\text{attack}}}{\text{anomaly}_{\text{benign}}}
\end{equation}

2. Capacity degradation: Captures the increase in system errors (\eg MAE, MSE, RMSE):
\begin{equation}
	AIS_{\text{C}} = \frac{\text{capacity}_{\text{attack}} - \text{capacity}_{\text{benign}}}{\text{capacity}_{\text{benign}}}
\end{equation}

3. Global AIS combines both metrics:
\begin{equation}
	AIS = \min ( \: \max \left( 0, AIS_{\text{A}} + AIS_{\text{C}}),\: 1 \: \right)
\end{equation}
A higher AIS indicates a stronger attack impact. An AIS of 0 denotes negligible effect, while AIS=1 corresponds to a worst-case degradation.

Finally, for backdoor attacks, we the \textit{Attack Success Rate (ASR)}, which measures the proportion of backdoored samples misclassified due to the attack:

\begin{equation}
	\text{ASR} = \frac{\sum_{i \in \text{backdoored indices}} \mathbb{I}(y_{\text{pred}}[i] \neq y_{\text{true}}[i])}{|\text{backdoored indices}|}
\end{equation}

where, \( y_{\text{true}}[i] \) is the original label of the \( i \)-th backdoored sample and \( y_{\text{pred}}[i] \) is its predicted label of the \( i \). \( \mathbb{I}(\cdot) \) is the indicator function, which returns 1 if the condition is true (\ie the prediction differs from the true label) and 0 otherwise. \textit{backdoored indices} refers to the indices of the backdoored samples in the test dataset. \( |\text{backdoored indices}| \) is the total number of backdoored test samples that normalize the metric.

\subsubsection{Comparison with C-DFL methods} 

The prior works on FL for EV battery management, summarized in Table~\ref{tab:comparison} rely on trusted servers or aggregators and primarily employ weighted variants of FedAvg. While such schemes can slightly improve convergence in benign settings, they remain vulnerable to poisoning attacks and lack robust Byzantine defenses. To provide a fair and comprehensive evaluation, we therefore define a set of baselines drawn from the general FL literature—covering both centralized and decentralized Byzantine-robust aggregation methods—and adapt them to the C-DFL setting. Specifically, we compare FLECA against the following:  FedAvg~\cite{mcmahan2017communication}, FedProx~\cite{wang2021field}, Multi-Krum~\cite{blanchard2017machine}, Norm Clipping and Weak-DP~\cite{sun2019can}, FLAME~\cite{Flame}, UBAR~\cite{guo2021byzantine}. The details of each method and their adaptation to C-DFL setting are provided in \textbf{Appendix~E.2}.
\color{black}

\subsubsection{Parameter settings}
We evaluate multiple C-DFL multitask models under a range of configurations by varying the total number of EVs $E = \sum_s \mathcal{E}_s$ and the group size $k = \mathcal{E}_s$. Specifically, we consider $E \in \{42, 70, 84, 126, 245\}$ and $k \in \{7, 10, 14, 21\}$. Unless stated otherwise, the default configuration is denoted as $\textit{Net-(42,7)}$. The adversarial setup is kept identical across all methods, attacks, and configurations: \textit{$\frac{1}{3}$ of the groups are fully malicious}, and within the remaining groups, \textit{$\frac{1}{3}$ of the EVs are Byzantine}. For all experiments, we set the DP parameters to $\sigma=0.005$ and $C=4.0$, and the aggregation parameters to $\beta \in [0.1, 0.3]$, $\kappa = 1.0$, and $minPts\in[2,3]$. The default training parameters are summarized in Table~\ref{tab:training_config}. A detailed ablation study analyzing sensitivity to these parameters is provided in
\textbf{Appendix~E.5}.

\subsubsection{Training configuration}
All experiments are conducted on three NVIDIA A6000 GPUs. Each experiment is repeated at least three times using different global random seeds (42, 70, and 84), and we report the mean and standard deviation (Std) of the results. The blockchain benchmarks are conducted on a cluster of two HPE ProLiant XL225n Gen10 Plus servers. Each server has two AMD EPYC 7713 64-Core 2GHz processors and 2x256 GB RAM. 
\color{black}

\begin{table}[t]
	\centering
	\caption{ Training Configuration.}
	\begin{tabular}{ll}
		\toprule
		\textbf{Parameter} & \textbf{Value} \\  \midrule
		%\textbf{Number of EVs} & 42, 70, 84, 126, 245 \\ 
		\textbf{Number of rounds} & 50 and 100 \\
		\textbf{Dirichlet $\alpha$} & 0.8 \\ 
		\textbf{Local epochs} & 20 \\ 
		\textbf{Batch size} & 32 \\ 
		\textbf{Learning rate} & 0.001 \\ 
		\textbf{Proximal regularization} & $\mu = 0.2$ \\ 
		\textbf{Dropout} & 0.3 \\ 
		\textbf{Early stopping} & Enabled (Patience: 10 rounds) \\ \bottomrule
	\end{tabular}
	\label{tab:training_config}
\end{table}

\begin{table}[th]
	\centering
	\caption{Performance comparison across anomaly and capacity tasks under benign settings using MultiTaskbiLSTM.}
	\setlength{\tabcolsep}{3pt}
	\resizebox{0.5\textwidth}{!}{
		\begin{tabular}{@{}ll>{\columncolor{blue!6}}lcc>{\columncolor{blue!6}}cccc@{}}
			\toprule
			\multirow{2}{*}{\textbf{Task}} &
			\multirow{2}{*}{\textbf{Metric}} &
			\multicolumn{3}{c}{\textbf{IID}} &
			\multicolumn{3}{c}{\textbf{Non-IID}} \\
			\cmidrule(lr){3-5} \cmidrule(lr){6-8}
			& & \textbf{FLECA} & \textbf{FedProx} & \textbf{FedAvg}
			& \textbf{FLECA} & \textbf{FedProx} & \textbf{FedAvg} \\
			\midrule
			
			\multirow{4}{*}{\textbf{Anomaly}}
			& Accuracy
			& 0.967$\pm$0.005 & 0.965$\pm$0.005 & 0.965$\pm$0.005
			& 0.964$\pm$0.010 & 0.965$\pm$0.010 & 0.968$\pm$0.010 \\
			& Recall
			& 0.930$\pm$0.010 & 0.925$\pm$0.010 & 0.927$\pm$0.010
			& 0.945$\pm$0.010 & 0.921$\pm$0.010 & 0.928$\pm$0.010 \\
			& Precision
			& 0.980$\pm$0.010 & 0.980$\pm$0.010 & 0.977$\pm$0.005
			& 0.957$\pm$0.020 & 0.983$\pm$0.020 & 0.984$\pm$0.020 \\
			& F1-Score
			& 0.954$\pm$0.005 & 0.952$\pm$0.005 & 0.951$\pm$0.005
			& 0.951$\pm$0.010 & 0.951$\pm$0.010 & 0.955$\pm$0.010 \\
			\cmidrule(lr){2-8}
			
			\multirow{3}{*}{\textbf{Capacity}}
			& MAE
			& 0.458$\pm$0.005 & 0.457$\pm$0.005 & 0.456$\pm$0.005
			& 0.465$\pm$0.010 & 0.459$\pm$0.010 & 0.462$\pm$0.010 \\
			& MSE
			& 0.665$\pm$0.005 & 0.669$\pm$0.005 & 0.671$\pm$0.005
			& 0.672$\pm$0.010 & 0.670$\pm$0.010 & 0.669$\pm$0.010 \\
			& RMSE
			& 0.815$\pm$0.005 & 0.818$\pm$0.005 & 0.819$\pm$0.005
			& 0.820$\pm$0.010 & 0.819$\pm$0.010 & 0.817$\pm$0.010 \\
			\bottomrule
		\end{tabular}
	}
	\label{tab:comparison_normal}
\end{table}

\begin{figure}[t]
	\centering
	
	\subfloat[IID]{
		\hspace{-0.2in}
		\includegraphics[scale=0.175]{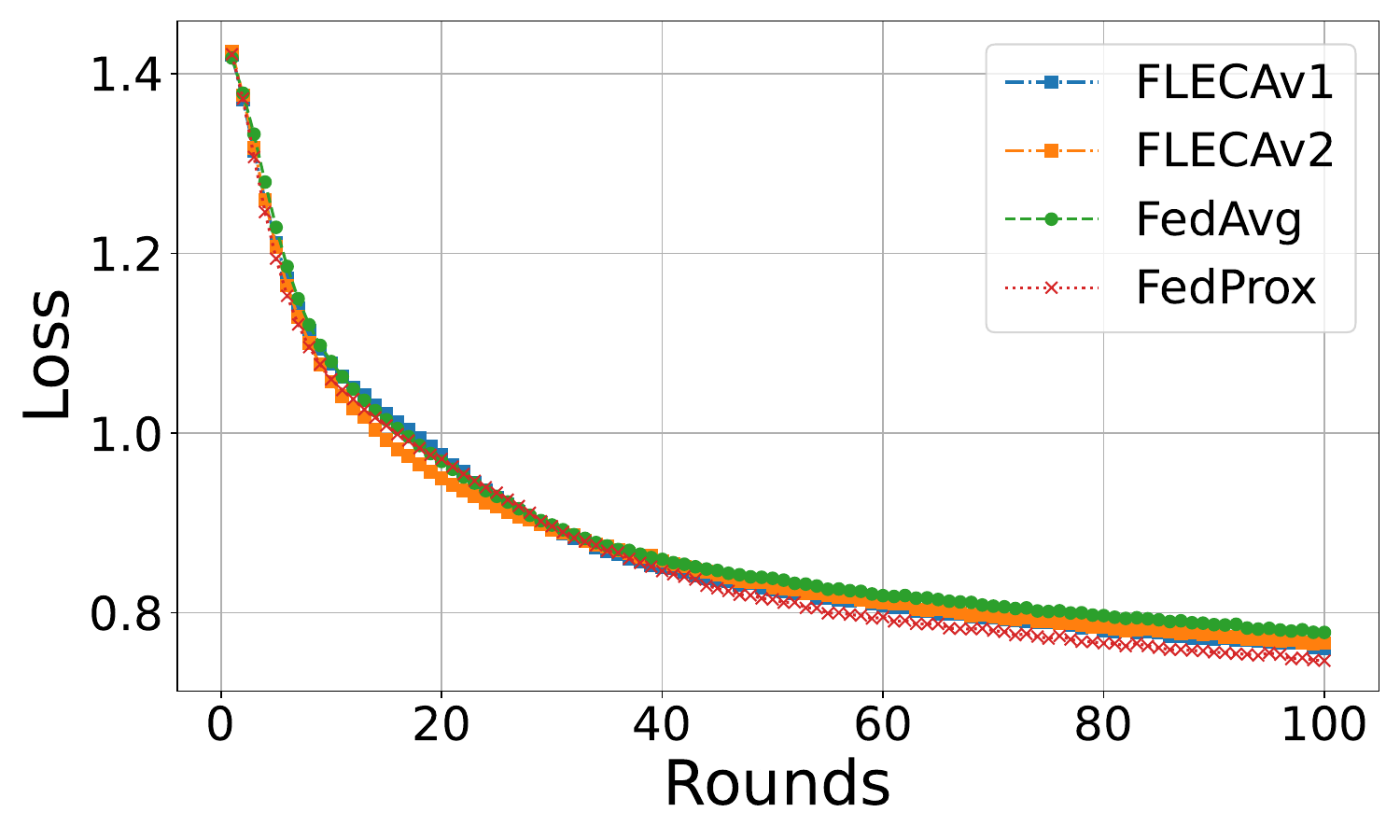}
		\label{fig15a}
	}
	\subfloat[Non-IID]{
		\hspace{-0.2in}
		\includegraphics[scale=0.175]{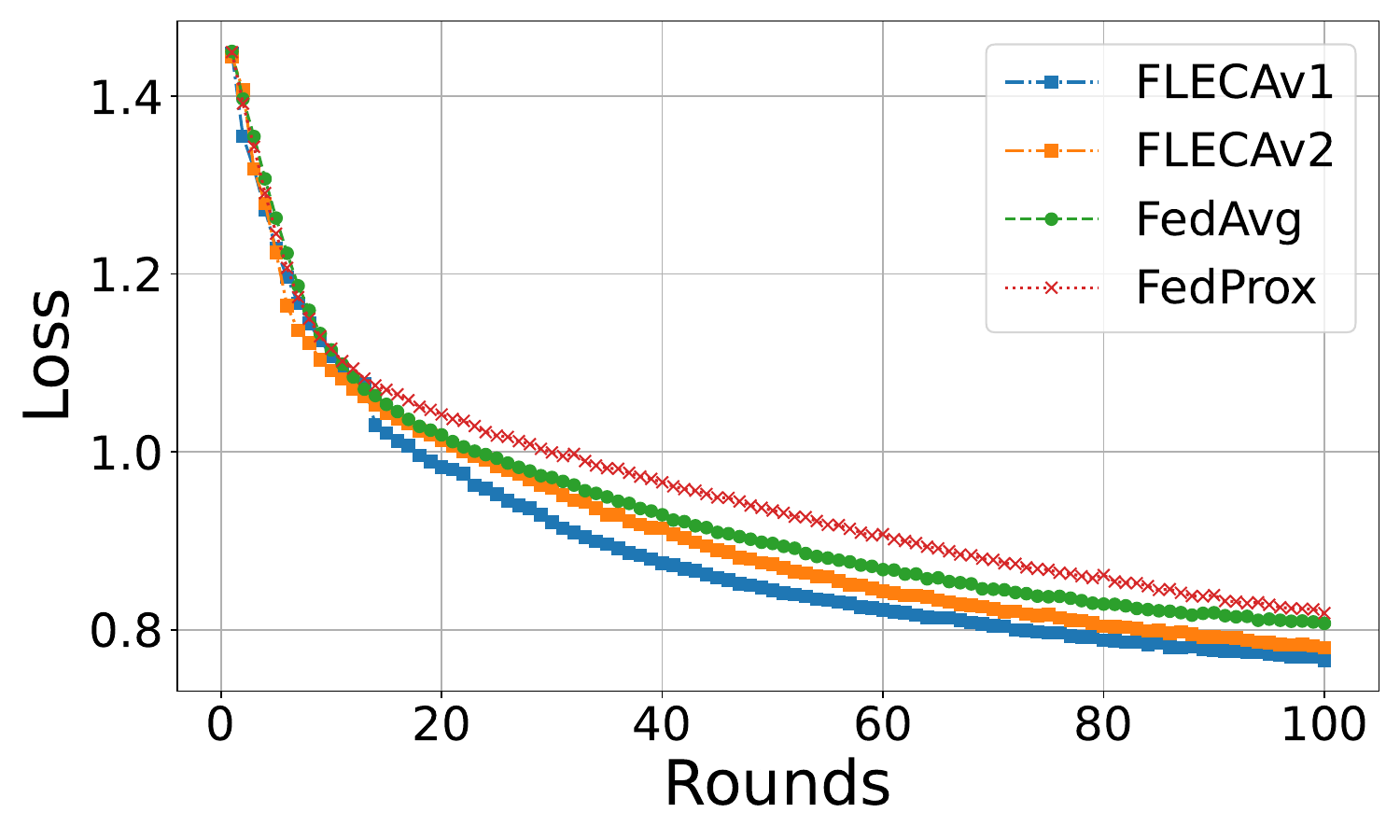}
		\label{fig15b}
	}
	\caption{Global loss (both tasks) evolution under normal conditions.}
	\label{fig15}
\end{figure}

\begin{table}[t]
	
	\centering
	\footnotesize
	\caption{Comparison of C-DFL (FLECA) with various FL systems under benign settings using biLSTM models.}
	\label{tab:cdfl_vs_all}
	\resizebox{\columnwidth}{!}{
		\begin{tabular}{lcccc}
			\toprule
			\textbf{Model} &
			\textbf{Accuracy} &
			\textbf{F1-Score} &
			\textbf{MAE} &
			\textbf{RMSE} \\
			
			\midrule
			ST-FL (FedProx) & 0.963$\pm$0.001 & 0.949$\pm$0.001 & 0.465$\pm$0.010 & 0.821$\pm$0.010 \\
			MT-FL (FedProx) & 0.965$\pm$0.000 & 0.952$\pm$0.000 & 0.458$\pm$0.001 & 0.814$\pm$0.000 \\
			SMT-FL (FedProx)      & 0.968$\pm$0.001 & 0.959$\pm$0.000 & 0.445$\pm$0.001 & 0.790$\pm$0.001 \\
			MT-DFL (FedProx) & 0.940$\pm$0.002 & 0.912$\pm$0.003 & 0.492$\pm$0.003 & 0.842$\pm$0.002 \\
			\rowcolor{blue!8}  MT-C-DFL (FLECAv1) & 0.967$\pm$0.005 & 0.956$\pm$0.005 & 0.459$\pm$0.002 & 0.816$\pm$0.005 \\
			\rowcolor{blue!8}  MT-C-DFL (FLECAv2) & 0.964$\pm$0.005 & 0.954$\pm$0.002 & 0.461$\pm$0.005 & 0.819$\pm$0.000 \\
			\bottomrule
	\end{tabular}}
	\noindent \textit{Abbrev. ST: Single Task, MT: Multi-Task, SMT: Split Multi-Task.}
\end{table}

\begin{table}[t]
	\centering
	\caption{ AIS evaluation under poisoning attacks and \textbf{IID} setting (mean $\pm$ std across 3 runs). Lower AIS is better. The benign FedProx baseline achieves an anomaly F1-score of $0.943 \pm 0.001$ and a capacity RMSE of $0.845 \pm 0.001$.}
	\label{tab:ais_poisoning}
	\resizebox{\columnwidth}{!}{
		\begin{tabular}{llccccc}
			\toprule
			\textbf{Attack} & \textbf{Method} &
			\textbf{Max.AIS} &
			\textbf{AIS$_A$} &
			\textbf{AIS$_C$} &
			\textbf{Anom. F1} &
			\textbf{Cap. RMSE} \\
			\midrule
			
			% ---------------- Gauss ----------------
			\multirow{9}{*}{Gauss}
			& FedAvg        & 1.000 & 0.345 & 1.000 & $0.617 \pm 0.089$ & $2.189 \pm 0.268$ \\
			& FedProx      & 1.000 & 0.387 & 0.906 & $0.578 \pm 0.076$ & $1.610 \pm 0.206$ \\
			& Norm-Clip    & 0.100 & 0.082 & 0.018 & $0.865 \pm 0.020$ & $0.859 \pm 0.001$ \\
			& Weak-DP      & 0.314 & 0.223 & 0.091 & $0.732 \pm 0.078$ & $0.921 \pm 0.027$ \\
			& Trimmed-Mean & 1.000 & 0.324 & 1.000 & $0.637 \pm 0.048$ & $2.355 \pm 0.807$ \\
			& Multi-Krum   & 0.000 & 0.000 & 0.000 & $0.946 \pm 0.003$ & $0.839 \pm 0.001$ \\
			& FLAME        & 0.029 & 0.026 & 0.003 & $0.918 \pm 0.003$ & $0.847 \pm 0.002$ \\
			& UBAR         & 0.000 & 0.000 & 0.000 & $0.943 \pm 0.001$ & $0.836 \pm 0.001$ \\
			\rowcolor{blue!8}  & \textbf{FLECA}        & 0.000 & 0.000 & 0.000 & $0.943 \pm 0.003$ & $0.838 \pm 0.002$ \\
			
			\midrule
			% ---------------- Krum ----------------
			\multirow{9}{*}{Krum}
			& FedAvg        & 1.000 & 0.426 & 1.000 & $0.541 \pm 0.000$ & $10.000 \pm 10.000$ \\
			& FedProx      & 1.000 & 0.426 & 1.000 & $0.541 \pm 0.000$ & $10.000 \pm 10.000$ \\
			& Norm-Clip    & 1.000 & 0.385 & 1.000 & $0.580 \pm 0.003$ & $10.000 \pm 0.642$ \\
			& Weak-DP      & 1.000 & 0.401 & 1.000 & $0.565 \pm 0.001$ & $10.000 \pm 2.350$ \\
			& Trimmed-Mean & 1.000 & 0.426 & 1.000 & $0.541 \pm 0.000$ & $10.000 \pm 10.000$ \\
			& Multi-Krum   & 0.001 & 0.001 & 0.000 & $0.941 \pm 0.000$ & $0.843 \pm 0.001$ \\
			& FLAME        & 0.004 & 0.004 & 0.000 & $0.939 \pm 0.008$ & $0.836 \pm 0.001$ \\
			& UBAR         & 0.023 & 0.018 & 0.005 & $0.926 \pm 0.002$ & $0.849 \pm 0.001$ \\
			\rowcolor{blue!8}  & \textbf{FLECA}        & 0.010 & 0.010 & 0.000 & $0.933 \pm 0.009$ & $0.837 \pm 0.002$ \\
			
			\midrule
			% ---------------- Trim ----------------
			\multirow{9}{*}{Trim}
			& FedAvg        & 0.848 & 0.696 & 0.152 & $0.286 \pm 0.265$ & $0.973 \pm 0.007$ \\
			& FedProx      & 0.586 & 0.425 & 0.161 & $0.542 \pm 0.002$ & $0.981 \pm 0.010$ \\
			& Norm-Clip    & 0.213 & 0.152 & 0.061 & $0.799 \pm 0.017$ & $0.896 \pm 0.002$ \\
			& Weak-DP      & 0.194 & 0.135 & 0.059 & $0.801 \pm 0.096$ & $0.891 \pm 0.008$ \\
			& Trimmed-Mean & 0.556 & 0.399 & 0.158 & $0.567 \pm 0.016$ & $0.978 \pm 0.004$ \\
			& Multi-Krum   & 0.000 & 0.000 & 0.000 & $0.943 \pm 0.001$ & $0.844 \pm 0.001$ \\
			& FLAME        & 0.012 & 0.009 & 0.003 & $0.934 \pm 0.003$ & $0.847 \pm 0.001$ \\
			& UBAR         & 0.046 & 0.046 & 0.000 & $0.899 \pm 0.002$ & $0.841 \pm 0.001$ \\
			\rowcolor{blue!8}  & \textbf{FLECA}        & 0.004 & 0.003 & 0.002 & $0.940 \pm 0.004$ & $0.846 \pm 0.001$ \\
			
			\midrule
			% ---------------- L-Flip ----------------
			\multirow{9}{*}{L-Flip}
			& FedAvg        & 0.399 & 0.399 & 0.000 & $0.567 \pm 0.001$ & $0.832 \pm 0.001$ \\
			& FedProx      & 0.406 & 0.406 & 0.000 & $0.560 \pm 0.000$ & $0.838 \pm 0.001$ \\
			& Norm-Clip    & 0.378 & 0.378 & 0.000 & $0.587 \pm 0.001$ & $0.842 \pm 0.001$ \\
			& Weak-DP      & 0.372 & 0.376 & 0.000 & $0.588 \pm 0.002$ & $0.840 \pm 0.003$ \\
			& Trimmed-Mean & 0.406 & 0.406 & 0.000 & $0.560 \pm 0.000$ & $0.838 \pm 0.001$ \\
			& Multi-Krum   & 0.426 & 0.426 & 0.000 & $0.541 \pm 0.000$ & $0.833 \pm 0.001$ \\
			& FLAME        & 0.301 & 0.301 & 0.000 & $0.658 \pm 0.133$ & $0.843 \pm 0.002$ \\
			& UBAR         & 0.014 & 0.014 & 0.000 & $0.929 \pm 0.003$ & $0.840 \pm 0.001$ \\
			\rowcolor{blue!8}  & \textbf{FLECA}        & 0.012 & 0.012 & 0.000 & $0.931 \pm 0.007$ & $0.840 \pm 0.002$ \\
			
			\midrule
			% ---------------- Feature ----------------
			\multirow{9}{*}{Feature}
			& FedAvg        & 0.807 & 0.037 & 0.770 & $0.908 \pm 0.001$ & $1.495 \pm 0.034$ \\
			& FedProx      & 1.000 & 0.043 & 1.000 & $0.902 \pm 0.003$ & $1.691 \pm 0.073$ \\
			& Norm-Clip    & 1.000 & 0.033 & 1.000 & $0.911 \pm 0.003$ & $2.705 \pm 0.049$ \\
			& Weak-DP      & 1.000 & 0.028 & 1.000 & $0.918 \pm 0.008$ & $2.526 \pm 0.053$ \\
			& Trimmed-Mean & 1.000 & 0.044 & 1.000 & $0.901 \pm 0.003$ & $1.692 \pm 0.078$ \\
			& Multi-Krum   & 0.000 & 0.000 & 0.000 & $0.952 \pm 0.001$ & $0.837 \pm 0.001$ \\
			& FLAME        & 0.018 & 0.018 & 0.000 & $0.925 \pm 0.003$ & $0.839 \pm 0.001$ \\
			& UBAR         & 0.010 & 0.010 & 0.000 & $0.933 \pm 0.001$ & $0.834 \pm 0.001$ \\
			\rowcolor{blue!8}  & \textbf{FLECA}        & 0.010 & 0.010 & 0.000 & $0.933 \pm 0.002$ & $0.837 \pm 0.002$ \\
			
			\midrule
			% ---------------- Adaptive ----------------
			\multirow{9}{*}{Adaptive}
			& FedAvg        & 1.000 & 0.032 & 1.000 & $0.912 \pm 0.008$ & $8.850 \pm 2.968$ \\
			& FedProx      & 1.000 & 0.032 & 1.000 & $0.912 \pm 0.008$ & $8.850 \pm 2.968$ \\
			& Norm-Clip    & 1.000 & 0.385 & 1.000 & $0.580 \pm 0.003$ & $10.000 \pm 0.642$ \\
			& Weak-DP      & 1.000 & 0.401 & 1.000 & $0.565 \pm 0.001$ & $10.000 \pm 2.350$ \\
			& Trimmed-Mean & 1.000 & 0.020 & 1.000 & $0.924 \pm 0.005$ & $5.714 \pm 1.680$ \\
			& Multi-Krum   & 0.006 & 0.006 & 0.000 & $0.937 \pm 0.002$ & $0.835 \pm 0.001$ \\
			& FLAME        & 0.008 & 0.008 & 0.000 & $0.935 \pm 0.001$ & $0.837 \pm 0.002$ \\
			& UBAR         & 0.008 & 0.008 & 0.000 & $0.935 \pm 0.001$ & $0.838 \pm 0.004$ \\
			\rowcolor{blue!8}  & \textbf{FLECA}        & 0.013 & 0.013 & 0.000 & $0.930 \pm 0.003$ & $0.838 \pm 0.001$ \\
			
			\bottomrule
	\end{tabular}}
	\newline
	\footnotesize \noindent\textit{Note. UBAR relies on a reference model and a local evaluation set for loss filtering, and can only be applied as a per-EV filtering mechanism. For a fair comparison, it additionally incorporates two components from FLECA: majority voting and inter-group robust clustering.}
	
\end{table}

\begin{table}[t]
	\centering
	
	\caption{ AIS evaluation under poisoning attacks in the \textbf{Non-IID} setting (mean $\pm$ std over 3 runs). Lower AIS is better. The benign FedProx baseline achieves an anomaly F1-score of $0.92 \pm 0.001$ and a capacity RMSE of $0.837 \pm 0.001$.}
	\label{tab:ais_poisoning_noniid}
	\resizebox{\columnwidth}{!}{
		\begin{tabular}{llccccc}
			\toprule
			\textbf{Attack} & \textbf{Method} &
			\textbf{Max.AIS} &
			\textbf{AIS$_A$} &
			\textbf{AIS$_C$} &
			\textbf{Anom. F1} &
			\textbf{Cap. RMSE} \\
			\midrule
			
			% ---------------- Gauss ----------------
			\multirow{9}{*}{Gauss}
			& FedAvg        & 1.000 & 0.356 & 0.705 & $0.596 \pm 0.070$ & $1.426 \pm 0.078$ \\
			& FedProx      & 1.000 & 0.334 & 1.000 & $0.616 \pm 0.060$ & $2.322 \pm 0.543$ \\
			& Norm-Clip    & 0.104 & 0.066 & 0.037 & $0.864 \pm 0.006$ & $0.868 \pm 0.003$ \\
			& Weak-DP      & 0.028 & 0.027 & 0.001 & $0.901 \pm 0.004$ & $0.837 \pm 0.001$ \\
			& Trimmed-Mean & 1.000 & 0.324 & 1.000 & $0.626 \pm 0.031$ & $2.116 \pm 0.406$ \\
			& Multi-Krum   & 0.043 & 0.027 & 0.016 & $0.901 \pm 0.005$ & $0.850 \pm 0.001$ \\
			& FLAME        & 0.069 & 0.040 & 0.028 & $0.888 \pm 0.027$ & $0.860 \pm 0.008$ \\
			& UBAR         & 0.142 & 0.087 & 0.055 & $0.846 \pm 0.003$ & $0.883 \pm 0.001$ \\
			\rowcolor{blue!8}  & \textbf{FLECA}        & 0.047 & 0.029 & 0.018 & $0.899 \pm 0.012$ & $0.852 \pm 0.007$ \\
			
			\midrule
			% ---------------- Krum ----------------
			\multirow{9}{*}{Krum}
			& FedAvg        & 1.000 & 0.416 & 1.000 & $0.541 \pm 0.000$ & $10.000 \pm 10.000$ \\
			& FedProx      & 1.000 & 0.416 & 1.000 & $0.541 \pm 0.000$ & $10.000 \pm 10.000$ \\
			& Norm-Clip    & 1.000 & 0.380 & 1.000 & $0.574 \pm 0.001$ & $10.000 \pm 1.037$ \\
			& Weak-DP      & 1.000 & 0.379 & 1.000 & $0.575 \pm 0.001$ & $10.000 \pm 8.961$ \\
			& Trimmed-Mean & 1.000 & 0.416 & 1.000 & $0.541 \pm 0.000$ & $10.000 \pm 10.000$ \\
			& Multi-Krum   & 0.050 & 0.035 & 0.015 & $0.894 \pm 0.003$ & $0.849 \pm 0.001$ \\
			& FLAME        & 0.030 & 0.022 & 0.009 & $0.906 \pm 0.012$ & $0.844 \pm 0.006$ \\
			& UBAR         & 0.106 & 0.058 & 0.048 & $0.872 \pm 0.003$ & $0.877 \pm 0.002$ \\
			\rowcolor{blue!8}  & \textbf{FLECA}        & 0.050 & 0.022 & 0.028 & $0.906 \pm 0.012$ & $0.860 \pm 0.007$ \\
			
			\midrule
			% ---------------- Trim ----------------
			\multirow{9}{*}{Trim}
			& FedAvg        & 0.536 & 0.368 & 0.168 & $0.585 \pm 0.019$ & $0.978 \pm 0.003$ \\
			& FedProx      & 0.456 & 0.267 & 0.190 & $0.679 \pm 0.028$ & $0.995 \pm 0.006$ \\
			& Norm-Clip    & 0.282 & 0.185 & 0.098 & $0.755 \pm 0.017$ & $0.918 \pm 0.001$ \\
			& Weak-DP      & 0.377 & 0.231 & 0.145 & $0.712 \pm 0.023$ & $0.958 \pm 0.018$ \\
			& Trimmed-Mean & 0.404 & 0.238 & 0.165 & $0.705 \pm 0.013$ & $0.975 \pm 0.001$ \\
			& Multi-Krum   & 0.096 & 0.051 & 0.045 & $0.878 \pm 0.008$ & $0.874 \pm 0.003$ \\
			& FLAME        & 0.142 & 0.078 & 0.065 & $0.854 \pm 0.020$ & $0.891 \pm 0.010$ \\
			& UBAR         & 0.097 & 0.057 & 0.040 & $0.873 \pm 0.002$ & $0.870 \pm 0.001$ \\
			\rowcolor{blue!8}  & \textbf{FLECA} & 0.108 & 0.082 & 0.027 & $0.850 \pm 0.005$ & $0.859 \pm 0.001$ \\

			\midrule
			% ---------------- L-Flip ----------------
			\multirow{9}{*}{L-Flip}
			& FedAvg        & 0.403 & 0.395 & 0.008 & $0.560 \pm 0.001$ & $0.843 \pm 0.001$ \\
			& FedProx      & 0.382 & 0.373 & 0.008 & $0.580 \pm 0.000$ & $0.844 \pm 0.001$ \\
			& Norm-Clip    & 0.388 & 0.370 & 0.017 & $0.583 \pm 0.000$ & $0.851 \pm 0.001$ \\
			& Weak-DP      & 0.382 & 0.366 & 0.016 & $0.587 \pm 0.004$ & $0.850 \pm 0.006$ \\
			& Trimmed-Mean & 0.382 & 0.374 & 0.008 & $0.580 \pm 0.000$ & $0.844 \pm 0.001$ \\
			& Multi-Krum   & 0.431 & 0.415 & 0.015 & $0.541 \pm 0.000$ & $0.850 \pm 0.001$ \\
			& FLAME        & 0.397 & 0.362 & 0.035 & $0.590 \pm 0.104$ & $0.866 \pm 0.004$ \\
			& UBAR         & 0.176 & 0.100 & 0.076 & $0.833 \pm 0.011$ & $0.900 \pm 0.001$ \\
			\rowcolor{blue!8}  & \textbf{FLECA}        & 0.093 & 0.059 & 0.034 & $0.871 \pm 0.002$ & $0.865 \pm 0.001$ \\
			
			\midrule
			% ---------------- Feature ----------------
			\multirow{9}{*}{Feature}
			& FedAvg        & 1.000 & 0.055 & 1.000 & $0.875 \pm 0.003$ & $1.715 \pm 0.203$ \\
			& FedProx      & 0.701 & 0.044 & 0.657 & $0.885 \pm 0.003$ & $1.386 \pm 0.045$ \\
			& Norm-Clip    & 1.000 & 0.108 & 1.000 & $0.825 \pm 0.005$ & $2.800 \pm 0.048$ \\
			& Weak-DP      & 1.000 & 0.110 & 1.000 & $0.822 \pm 0.007$ & $2.805 \pm 0.076$ \\
			& Trimmed-Mean & 0.699 & 0.043 & 0.656 & $0.886 \pm 0.003$ & $1.385 \pm 0.046$ \\
			& Multi-Krum  & 0.075 & 0.040 & 0.035 & $0.888\pm0.003$ & $0.866\pm0.003$ \\
			& FLAME       & 0.199 & 0.124 & 0.076 & $0.811\pm0.012$ & $0.900\pm0.005$ \\
			& UBAR         & 0.093 & 0.063 & 0.030 & $0.868 \pm 0.004$ & $0.862 \pm 0.001$ \\
			\rowcolor{blue!8}  & \textbf{FLECA}        & 0.080 & 0.046 & 0.034 & $0.883 \pm 0.006$ & $0.865 \pm 0.001$ \\
			
			\midrule
			% ---------------- Adaptive ----------------
			\multirow{9}{*}{Adaptive}
			& FedAvg        & 1.000 & 0.042 & 1.000 & $0.887 \pm 0.007$ & $8.365 \pm 2.747$ \\
			& FedProx      & 1.000 & 0.042 & 1.000 & $0.887 \pm 0.007$ & $8.365 \pm 2.747$ \\
			& Norm-Clip    & 1.000 & 0.380 & 1.000 & $0.574 \pm 0.001$ & $10.000 \pm 1.037$ \\
			& Weak-DP      & 1.000 & 0.379 & 1.000 & $0.575 \pm 0.001$ & $10.000 \pm 8.961$ \\
			& Trimmed-Mean & 1.000 & 0.022 & 1.000 & $0.906 \pm 0.006$ & $5.095 \pm 1.520$ \\
			& Multi-Krum   & 0.431 & 0.002 & 0.415 & $0.850 \pm 0.001$ & $0.541 \pm 0.000$ \\
			& FLAME        & 0.397 & 0.109 & 0.363 & $0.866 \pm 0.004$ & $0.590 \pm 0.104$ \\
			& UBAR         & 0.015 & 0.014 & 0.010 & $0.917 \pm 0.009$ & $0.841 \pm 0.005$ \\
			\rowcolor{blue!8}  & \textbf{FLECA}        & 0.019 & 0.005 & 0.014 & $0.913 \pm 0.004$ & $0.842 \pm 0.001$ \\
			
			\bottomrule
	\end{tabular}}
\end{table}

\subsection{Experimental Results}

\subsubsection{Benign performance}
We first evaluate FLECA in fully benign settings, where all EVs and groups behave correctly. As shown in Table~\ref{tab:comparison_normal} and Fig.~\ref{fig15}, FLECA achieves performance on par with standard FL baselines (FedAvg, FedProx) across both anomaly detection and capacity estimation tasks, under IID and Non-IID data distributions and multiple architectures (see \textbf{Appendix~E.3}).  
For instance, under IID anomaly detection, FLECA attains an accuracy of $0.967$, slightly exceeding FedAvg and FedProx ($0.965$).

Moreover, Table~\ref{tab:cdfl_vs_all} shows that FLECA matches or outperforms a broad range of single-task and multi-task FL baselines. In particular, it achieves an F1-score of $0.951$, comparable to ST-FL, MT-FL, and Split MT-FL, while outperforming graph-based MT-DFL methods that suffer from neighborhood-level convergence limitations.  
Overall, these results confirm that FLECA’s filtering mechanisms do not degrade benign performance and can even provide mild regularization benefits.

\subsubsection{Robustness to poisoning attacks}
We evaluate robustness under poisoning attacks with a strong adversarial setting: one-third of the groups are fully malicious, and one-third of EVs are adversarial within the remaining groups. Results are reported for IID and Non-IID (Dirichlet $\alpha=0.8$) data.

Tables~\ref{tab:ais_poisoning} and~\ref{tab:ais_poisoning_noniid} report Max.AIS. Across all attacks and distributions, FLECA consistently achieves among the lowest AIS values, indicating strong preservation of both tasks.

\textit{IID setting.}  
Under IID data, FLECA maintains near-zero Max.AIS for most poisoning attacks, including \emph{Gauss}, \emph{Krum}, \emph{Trim}, \emph{Feature}, and \emph{Adaptive}. In these cases, anomaly F1-scores remain close to the benign FedProx baseline ($\approx 0.94$) and capacity RMSE stays around $0.84$.  
In contrast, FedAvg, FedProx, and Trimmed Mean frequently collapse (Max.AIS $=1$). While Multi-Krum and FLAME perform well in some IID scenarios, FLECA matches or exceeds their robustness with lower variance and more stable convergence.  
Notably, under \emph{Label-Flipping} and \emph{Adaptive} attacks—where most baselines fail—FLECA reduces AIS to $\approx 0.01$ and restores anomaly F1-scores above $0.93$.

\textit{Non-IID setting.}  
Under Non-IID data, all methods experience degraded performance; nevertheless, FLECA remains consistently among the best defenses. For \emph{Gauss} and \emph{Krum} attacks, FLECA maintains AIS below $0.05$, outperforming classical baselines and remaining competitive with Multi-Krum and FLAME.  
For harder attacks such as \emph{Adaptive} and \emph{L-Flip}, FLECA still outperforms UBAR (e.g., Max.AIS $0.093$ vs.\ $0.176$), despite not relying on local evaluation signals (\ie local loss in UBAR). This demonstrates FLECA’s robustness to the combined effects of adversarial behavior and statistical heterogeneity.

\begin{table}[t]
	\centering
	
	\caption{ Maximum Attack Success Rate (Max.ASR) under backdoor attacks in \textbf{IID} and \textbf{Non-IID} settings. Lower ASR indicates stronger backdoor robustness.}
	\label{tab:asr_backdoor_merged}
	\resizebox{\columnwidth}{!}{
		\begin{tabular}{llcc}
			\toprule
			\textbf{Attack} & \textbf{Method} & \textbf{Max.ASR (IID)} & \textbf{Max.ASR (Non-IID)} \\
			\midrule
			\multirow{9}{*}{BadNets}
			& FedAvg        & 0.2719 & 0.2976 \\
			& FedProx       & 0.2719 & 0.2976 \\
			& Norm-Clip     & 0.2719 & 0.2960 \\
			& Weak-DP       & 0.3418 & 0.3708 \\
			& Trimmed-Mean  & 0.2727 & 0.3054 \\
			& Multi-Krum    & 0.4113 & 0.4256 \\
			& FLAME         & 0.3794 & 0.3442 \\
			& UBAR          & 0.1177 & 0.1609 \\
			\rowcolor{blue!8}  & \textbf{FLECA}        & 0.1235 & 0.1762 \\
			\midrule
			\multirow{9}{*}{Scaling}
			& FedAvg        & 0.7167 & 0.8136 \\
			& FedProx       & 0.7167 & 0.8111 \\
			& Norm-Clip     & 0.3238 & 0.4105 \\
			& Weak-DP       & 0.3765 & 0.1590 \\
			& Trimmed-Mean  & 0.7011 & 0.8029 \\
			& Multi-Krum    & 0.1190 & 0.1337 \\
			& FLAME         & 0.1026 & 0.1128 \\
			& UBAR          & 0.1177 & 0.1611 \\
			\rowcolor{blue!8}  & \textbf{FLECA}        & 0.1112 & 0.0981 \\
			\midrule
			\multirow{9}{*}{Neurotoxin}
			& FedAvg        & 0.7469 & 0.9305 \\
			& FedProx       & 0.7469 & 0.9256 \\
			& Norm-Clip     & 0.3213 & 0.4571 \\
			& Weak-DP       & 0.3921 & 0.4456 \\
			& Trimmed-Mean  & 0.6321 & 0.7584 \\
			& Multi-Krum    & 0.1190 & 0.2657 \\
			& FLAME         & 0.1026 & 0.8115 \\
			& UBAR          & 0.1177 & 0.0904 \\
			\rowcolor{blue!8}  & \textbf{FLECA}        & 0.1094 & 0.0948 \\
			\bottomrule
	\end{tabular}}
\end{table}

\subsubsection{Backdoor robustness}
Table~\ref{tab:asr_backdoor_merged} reports Max.ASR under representative backdoor attacks. Across IID and Non-IID settings, FLECA consistently achieves low ASR, comparable to or better than UBAR and significantly outperforming smoothing defenses (Norm-Clip, Weak-DP) and robust aggregators (Multi-Krum, Trimmed Mean).  
Specifically, FLECA limits ASR to approximately $0.11$--$0.12$ under IID and $0.09$--$0.18$ under Non-IID data, confirming strong resistance to backdoor injection even under heterogeneous distributions.

\subsubsection{Impact of malicious group ratio}
\begin{figure*}[th]
	\centering
	\includegraphics[width=1.0\linewidth]{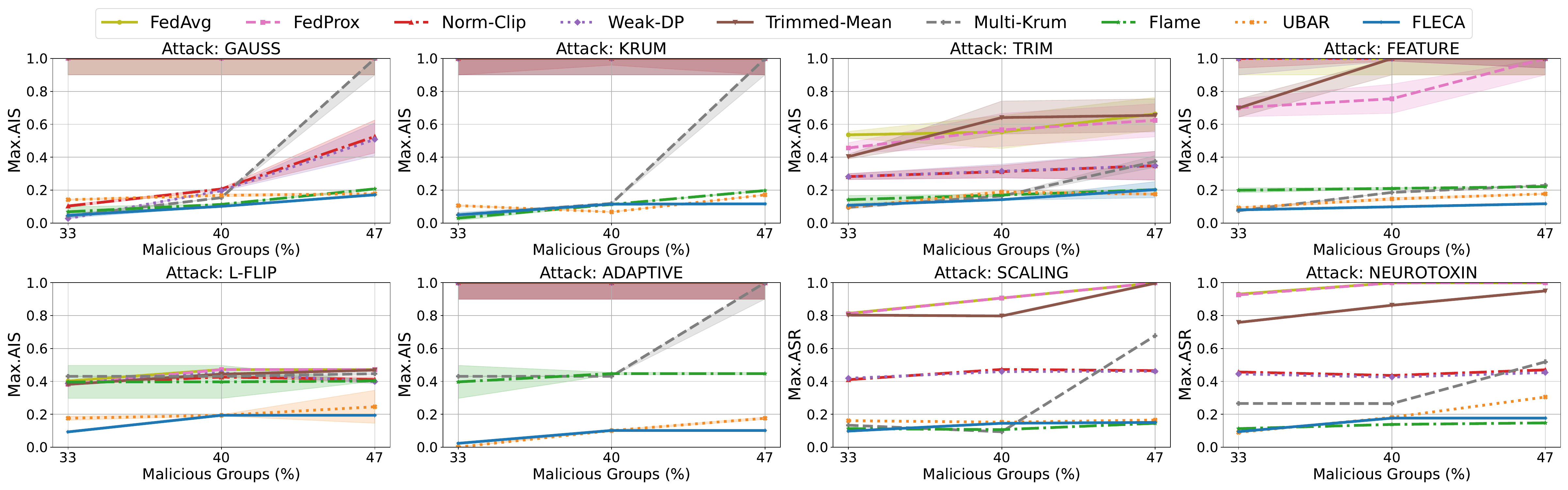}
	\caption{ Impact of the proportion of malicious groups with Non-IID ($\alpha =0.8$): AIS $=f($F1-Score, RMSE$)$.}
	\label{fig9}
\end{figure*}

Fig.~\ref{fig9} analyzes robustness as the proportion of malicious groups increases. FLECA remains effective even when more than $47\%$ of groups are malicious, while most baselines degrade rapidly. Among competitors, UBAR is the strongest baseline but relies on costly local evaluation signals.  
Smoothing defenses and lightweight robust aggregators fail under high adversarial ratios or adaptive attacks, whereas FLECA consistently suppresses poisoning and maintains Max.ASR below $0.2$ across all settings.

\begin{figure}[th]
	\centering
	\includegraphics[width=1.0\linewidth]{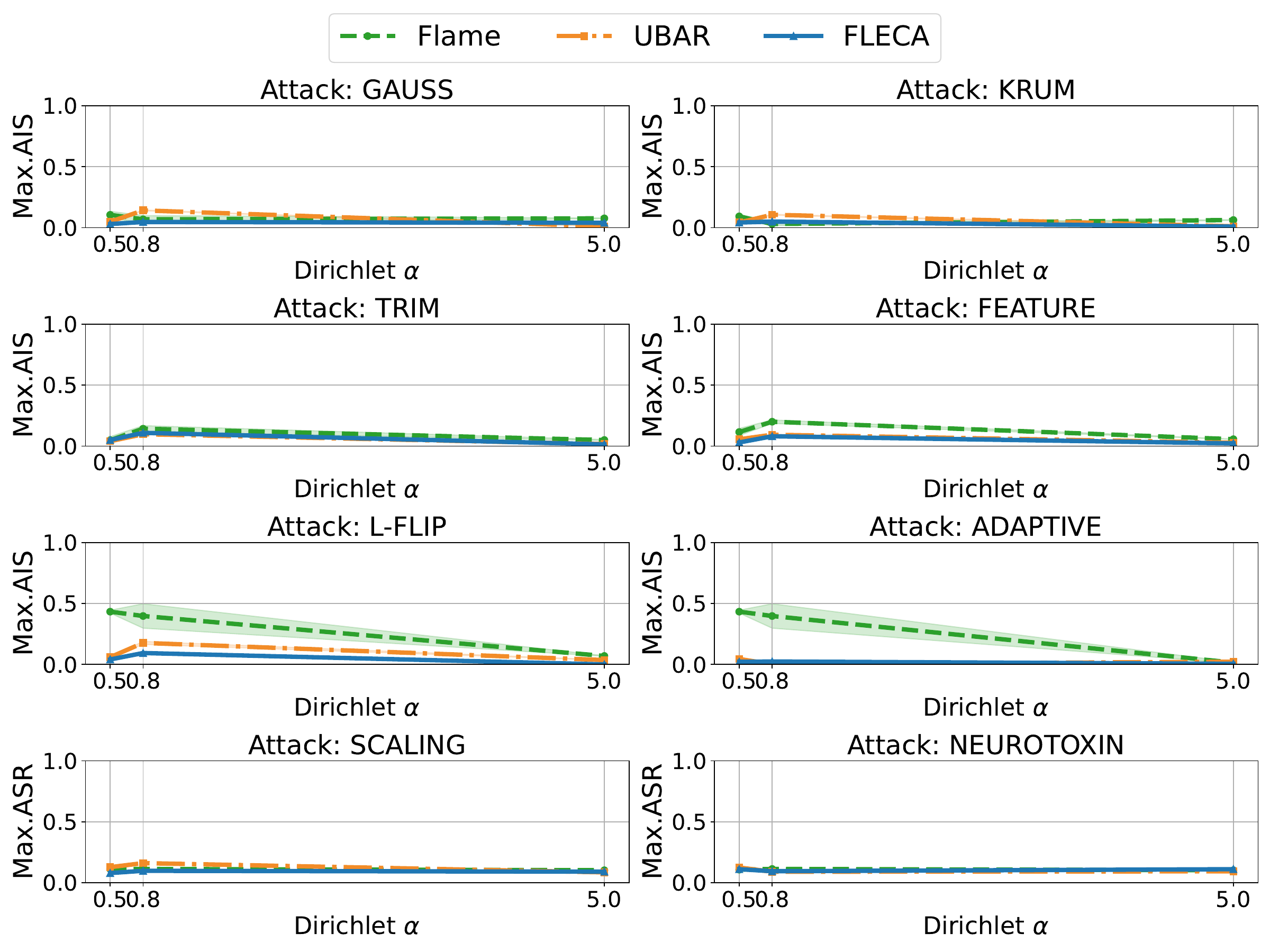}
	\caption{ Impact of Non-IIDness under attacks with \( \frac{1}{3} \) of malicious groups and \( \frac{1}{3} \) of malicious EVs within the remaining groups: AIS $=f($F1-Score, RMSE$)$.}
	\label{fig10}
\end{figure}

\subsubsection{Impact of data heterogeneity}
Fig.~\ref{fig10} shows that FLECA remains robust under increasing Non-IIDness, successfully separating benign and malicious updates even for highly skewed data ($\alpha=0.5$). This confirms that the proposed filtering mechanism is resilient to strong statistical heterogeneity.

\begin{figure}[th]
	\centering
	\includegraphics[width=0.95\linewidth]{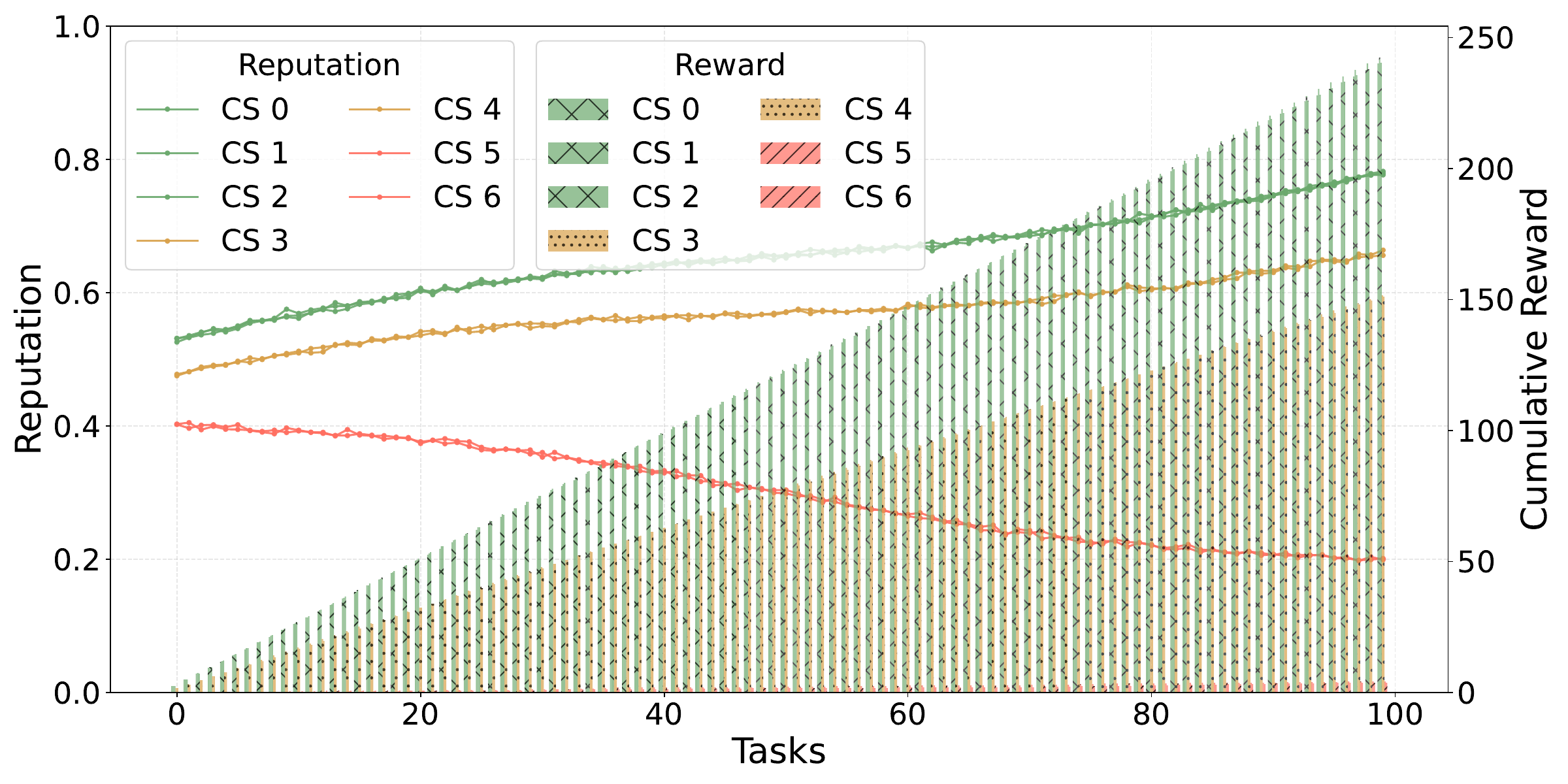}
	\caption{ Changes in CSs reputation $R$ and cumulative rewards $r$ (bars) over 10 training tasks. Reputation is updated using : 
		\(
		\text{R}_{\text{new}} = (1 - 0.1)\,\text{R}_{\text{old}} + 0.1\, e^{-5.0 e^{-0.5 S_s}} 
		\)}
	\label{fig14}
\end{figure}

\subsubsection{Incentive effectiveness}
We evaluate the hybrid incentive mechanism over ten training tasks involving seven CSs with heterogeneous behaviors. As shown in Fig.~\ref{fig14}, reliable CSs exhibit steadily increasing reputations and receive the highest cumulative rewards, while intermittent contributors earn moderate rewards. Malicious CSs experience continuous reputation decay and receive no rewards, as their poisoned updates are consistently filtered.  
These results demonstrate effective alignment between contribution quality, trust, and incentives.

\subsubsection{On-chain performance}

\begin{table}[th]
	\centering
	\caption{Deployment, invocation, and E2E gas costs for a 100-round learning task.}
	\label{tab:gascost_e2e}
	\resizebox{1.0\columnwidth}{!}{ 
		\begin{tabular}{lcccc}
			\toprule
			\textbf{Contract/Function}  & \textbf{Gas used} & \textbf{E2E Gas Pure BFL} & \textbf{E2E Gas ABC-DFL} \\   
			\midrule
			\textbf{ASC}               & 803,599      & 803,599       & 803,599 \\     
			\textbf{MSC}               & 5,898,853    & 5,898,853      & 5,898,853  \\     
			\texttt{registerMP}        & 92,267       & 92,267          & 92,267   \\  
			\texttt{publishModel}       & 197,677      & 19.77 M         & 19.77 M  \\   
			\texttt{register[CS/EV]}   & 114,462      & 114.5 B         & 11.45 B  \\     
			\texttt{join[CS/EV]Model}  & 97,523       & 97.5 B          & 9.75 B  \\     
			\texttt{submitIM}           & 107,788      & 107.8 B         & 10.78 B  \\     
			\texttt{submitGM}           & 102,753      & 10.28 M         & 10.28 M  \\     
			\texttt{update[CS/EV]Scores} & 124,501   & 124.5 B         & 12.45 B \\     
			\texttt{distributeReward}   & 119,295      & 119.3 B         & 11.93 B  \\     
			
			\bottomrule
		\end{tabular}
	}
\end{table}

\begin{figure}[th]
	\centering
	\includegraphics[width=0.95\linewidth]{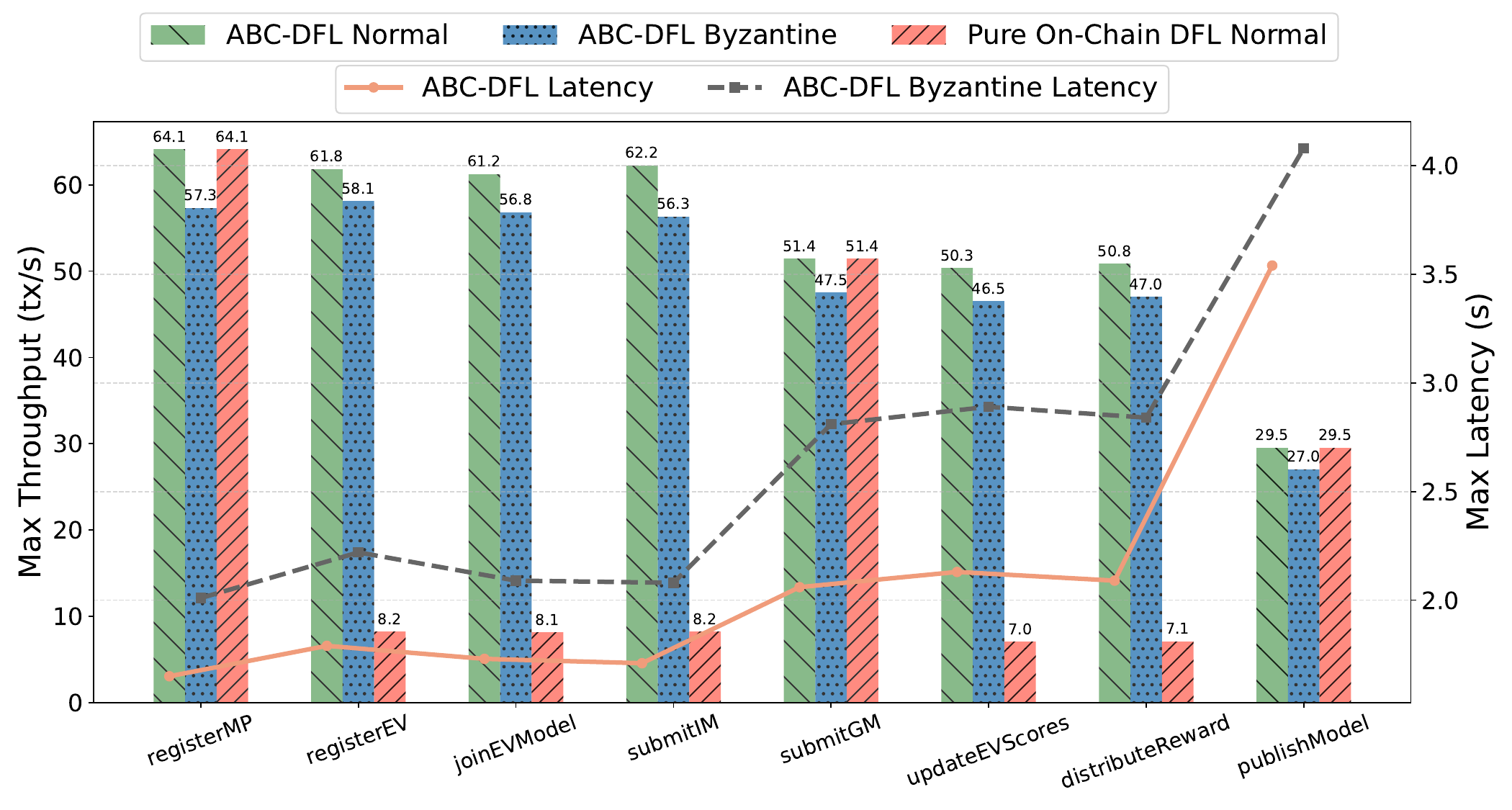}
	\caption{ Throughput and Latency of $ABC$-DFL with a workload of 10000 EVs (group size $k$=10).}
	\label{fig11}
\end{figure}

We evaluate the on-chain performance of $ABC$-DFL using Hyperledger Caliper\footnote{\href{https://github.com/hyperledger/caliper-benchmarks}{https://github.com/hyperledger/caliper-benchmarks}} on a local EVM-based network powered by Hyperledger Besu\footnote{\href{https://besu.hyperledger.org}{https://besu.hyperledger.org}} with Dynamic QBFT consensus. The transaction sending rate varies from 10 to 500 tx/s under a fixed configuration (block time = 1s, $\mathcal{V}=18$, 1000 CSs). Gas costs are reported in Table~\ref{tab:gascost_e2e}, while average latency and throughput under normal and Byzantine (3 validators) conditions are shown in Fig.~\ref{fig11}.

Under normal operation, lightweight functions such as \texttt{registerMP} and \texttt{submitGM} achieve the highest throughput (64.1 and 62.2 tx/s), whereas more computation-heavy functions (\texttt{publishModel}, \texttt{updateCSScores}, \texttt{distributeReward}) exhibit lower throughput. Latency remains bounded across all functions (1.65--3.54s).

Byzantine conditions induce a moderate throughput degradation and increased latency due to additional validation steps; however, the system sustains an average throughput of $\approx 50$ tx/s. Compared to a pure BFL design (Table~\ref{tab:gascost_e2e}) where each EV commits local updates on-chain, $ABC$-DFL aggregates updates at the CS level and stores only compact artifacts (hashes and aggregated signatures). This reduces the number of blockchain transactions by approximately a factor of $k$, substantially reducing gas costs and improving scalability. Finally, since $ABC$-DFL is EVM-compatible, Layer-2 solutions (e.g., zk-Rollups) can further reduce gas costs by an order of magnitude and scale throughput beyond 1000 tx/s~\cite{autodfl}.

\subsubsection{E2E overhead}
\label{sec:consensus_oracle_overhead}

We quantify the end-to-end (E2E) latency overhead introduced by blockchain consensus and oracle layers in $ABC$-DFL under realistic vehicular conditions. Detailed measurements are provided in \textbf{Appendix~E.4}.

\textit{Consensus.} Dynamic QBFT consistently outperforms PBFT and static QBFT, achieving low latency (0.9--3.2s) and stable liveness under intermittent connectivity and validator churn.

\textit{Oracle.} A consortium-operated edge oracle incurs 1.1--3.0s per global round for aggregation, evaluation, and callbacks—significantly lower than public oracle deployments and compatible with vehicular FL dynamics.

\textit{E2E rounds.} Combining local training, EV exchanges, oracle execution, and on-chain processing, a full $ABC$-DFL round completes in 4.70--8.60s, compared to 2.90--5.20s for centralized FL. While decentralization introduces bounded coordination overhead, it provides auditability, accountability, and fault tolerance.

\subsection{Discussion}
\label{sec:discussion}

Based on our experimental and ablation studies, we provide key insights into the behavior of FLECA and practical deployment of the underlying ABC-DFL framework:
\begin{itemize}
	\item \textit{Governance and identity management:} The system assumes an open yet permissioned model for CSs and MPs registration to Sybil and reputation reset attacks.
	\item \textit{Task-level scalability:} Only a subset of EVs participates per task, reducing network and computational load.
	\item \textit{Known failure modes:} Exceeding the tolerated Byzantine rates or extreme EV churn can destabilize convergence.
	\item \textit{Parameter recommendations:} Reasonable defaults are $k \in [7,28]$ and $E \in [42, 126]$, intra- and inter-group \textit{minPts} $\approx 3$, $\beta \in [0.1, 0.3] $ $\kappa \in [0.5,1.0]$, DP clipping $C=4.0$, and $\sigma \in [0.005,0.01]$.
\end{itemize}
FLECA currently operates on explicit update exchanges with controlled noise to enable filtering and verification, secure robust aggregation is a potential future research direction to strengthen data privacy. % by hiding individual EV contributions, provided that lightweight compatibility mechanisms are introduced to preserve the framework’s layered defense and validation pipeline.

\color{black}

\section{Conclusion} \label{sec:conclusion}
\quad In this paper, we presented $ABC$-DFL, A byzantine-resilient clustered decentralized federated learning (C-DFL) framework for battery intelligence in connected EVs and EV-powered ITS. By introducing a novel robust aggregation protocol, FLECA, built on top of an open-permissioned blockchain and a decentralized oracle network, we effectively mitigate the vulnerabilities associated with centralized aggregation protocols, ensuring security and efficiency. The proposed FLECA demonstrated strong resilience against poisoning attacks, with performance comparable to state-of-the-art methods in non-Byzantine environments. Extensive experiments validated the efficiency and robustness of $ABC$-DFL, positioning it as a practical, secure, and efficient solution for next-generation EV data management systems. Finally, this work also paves the way for further advancements in Clustred-DFL for critical applications in the transportation and energy sectors.

\section*{Acknowledgments}
This work is supported by the OPEVA project that has received funding within the Chips Joint Undertaking (Chips JU) from the EU’s Horizon Europe Programme and the National Authorities (France, Czechia, Italy, Portugal, Turkey, Switzerland), under grant agreement 101097267. BPI funds the project in France under the France 2030 program on ``Embedded AI''. Views and opinions expressed are however those of the author(s) only and do not necessarily reflect those of the EU or Chips JU. Neither the EU nor the granting authority can be held responsible for them.

%\newpage
\bibliographystyle{IEEEtran}
\bibliography{ref}

\end{document}